\documentclass[twocolumn,twoside,trackchanges]{aastex701}

\usepackage{appendix}
\usepackage{tabularx}
\usepackage{amsmath, amssymb}
\usepackage{appendix}
\usepackage{hyperref}
\usepackage{booktabs}
\usepackage{longtable}

\begin{document}

\title{Berkeley 32: A Metal-poor and Dynamically Evolved Open Cluster with Evidence of Radial Migration}

\correspondingauthor{D. Bisht}
\email{devendrabisht297@gmail.com}  

\author[orcid=0000-0001-7940-3731, sname=Çınar, gname=Deniz Cennet]{D. C. Çınar}\thanks{E-mail: denizcennetcinar@gmail.com}
\affiliation{Programme of Astronomy and Space Sciences, Institute of Graduate Studies in Science, Istanbul University, Istanbul, 34116, Turkey}
\email{denizcennetcinar@gmail.com}

\author[orcid=0000-0002-8988-8434, sname=Bisht, gname=D.]{D. Bisht}
\affiliation{Indian Centre For Space Physics
466, Barakhola, Singabari road, Netai Nagar, Kolkata, West Bengal, 700099}
\email{devendrabisht297@gmail.com}

\author[orcid=0000-0001-7359-3300, sname=Jiang, gname=D.]{Ing-Guey Jiang}\thanks{E-mail: jiang@phys.nthu.edu.tw}
\affiliation{Department of Physics and Institute of Astronomy, National Tsing-Hua University, Hsinchu 30013, Taiwan}
\email{jiang@phys.nthu.edu.tw}

\author[orcid=0000-0002-2298-4026, sname=Elsanhoury, gname=W. H.]{W. H. Elsanhoury}
\affiliation{Department of Physics, College of Science, Northern Border University, Arar, Saudi Arabia} \email{elsanhoury@nbu.edu.sa}

\author[sname=Belwal, gname=K.]{K. Belwal}
\affiliation{Indian Centre For Space Physics
466, Barakhola, Singabari road, Netai Nagar, Kolkata, West Bengal, 700099}
\email{kuldeepbelwal1997@gmail.com}

\author[0000-0003-3510-1509]{Sel\c{c}uk Bilir} \affiliation{Istanbul University, Faculty of Science, Department of Astronomy and Space Sciences, 34119, Beyaz\i t, Istanbul, Turkey} \email{sbilir@istanbul.edu.tr}

\begin{abstract}
We present a comprehensive chemo-dynamical analysis of the old, metal-poor open cluster Berkeley~32 based on Gaia DR3 astrometry and Gaia-ESO Survey DR5.1 spectroscopy. Cluster membership is determined using a Gaussian Mixture Model applied to proper-motion components and trigonometric parallaxes. Isochrone fitting yields an age of $4.9\pm 0.5$~Gyr, a heliocentric distance of $3325$~pc, and an extinction of $A_{\rm V}=0.38\pm0.12$~mag. Spectroscopic member stars exhibit a mean metallicity of $[\rm Fe/H]=-0.39\pm 0.02$~dex, near-solar $\alpha$-element abundances, and a weighted mean radial velocity of $V_{\rm rad} = 106.26 \pm 0.03$~km~s$^{-1}$. The [Y/Mg] chemical clock yields an age of $4.73 \pm 2.39$~Gyr, consistent with the isochrone estimate. Orbital integration indicates a moderately eccentric orbit ($e=0.268\pm 0.004$) with a guiding radius of $R_{\rm g} = 8.82$~kpc. The inferred chemical birth radius, $R_{\rm b} = 9.82$~kpc, together with $\Delta R\approx -1$~kpc, suggests moderate inward radial migration, while the offsets among $R_{\rm b}$, $R_{\rm g}$, and $R_{\rm GC}$ are consistent with both churning and blurring processes. A photometric analysis identifies a binary fraction of $f_{\rm b}=0.449\pm 0.017$ for systems with mass ratios $q\geq 0.5$, implying a substantial unresolved binary population. Radial cumulative distribution functions further reveal significant mass segregation, with evolved stars more centrally concentrated than main-sequence stars. These results indicate that Berkeley~32 is a dynamically evolved old-disk cluster whose present-day structure and orbit preserve signatures of both internal dynamical evolution and radial migration within the Galactic disk.

\end{abstract}


\section{Introduction}
\label{sec:intro}

The Milky Way disk has been shaped over billions of years through star formation, metal enrichment, and the dynamical rearrangement of its stellar populations. Understanding how the chemical composition of disk material evolves with time and the Galactocentric radius is a central goal of Galactic archaeology \citep{Freeman2002, BlandHawthorn2016}. The metallicity gradient, the gradual decrease in iron abundance with increasing Galactocentric distance \citep{Magrini2009, Onal2016}, reflects the integrated star formation history, the balance between infall of pristine gas and enrichment by stellar ejecta, and the radial transport of metals via large-scale Galactic flows \citep{Chiappini2001, Schoenrich2009, Minchev2013, Minchev2014}. Chemical evolution models capture these processes using nucleosynthesis yields, supernova rates, and gas infall timescales \citep{Romano2010, Kobayashi2020, Palla2022}, yet observational constraints at intermediate Galactocentric radii ($R_\mathrm{GC} \sim 7$--9~kpc) remain sparse compared to the solar neighbourhood, and the temporal evolution of the gradient is still debated \citep{Magrini2017, Spina2022, Myers2022, Magrini2023}.

Among the most informative tracers of Galactic disk chemical evolution are open clusters (OCs), whose member stars share a common age, distance, and chemical composition, enabling reliable measurements of metallicity and elemental abundances \citep{Friel1995}. Large spectroscopic surveys such as the Gaia-ESO Survey \citep[GES;][]{Gilmore2022, Randich2022}, GALAH \citep{Buder2021}, and APOGEE \citep{Majewski2017} have produced homogeneous abundance catalogs for about a hundred OCs, revealing the Galactic radial metallicity gradient and its evolution with cluster age \citep{Donor2020, Spina2022}. Gaia DR3 \citep{GaiaCollaboration2023} provides precise astrometry and radial velocities for hundreds of millions of stars, enabling accurate six-dimensional phase-space studies. Building on these data, \citet{Hunt2024} compiled a catalog of approximately 7,000 OC candidates with homogeneous membership lists and mean astrometric parameters, providing a foundation for systematic orbital studies.

A major challenge, however, comes from radial migration: gravitational perturbations from spiral arms and the Galactic bar can scatter clusters and their stars away from their birth radii ($R_{\rm b}$) \citep{Sellwood2002, Roskar2008}. Consequently, clusters observed today at a given $R_\mathrm{GC}$ may have originated at quite different Galactocentric distances, meaning that the observed metallicity--radius relation mixes the gradient at formation with later dynamical rearrangement of the disk \citep{Anders2017, Quillen2018, Frankel2020}. Reconstructing $R_{\rm b}$ through backward orbital integration is therefore crucial to separate the imprint of chemical evolution from the effects of radial migration \citep{Minchev2018, Feltzing2020}. Despite recent advances, observational constraints on the temporal evolution of Galactic metallicity at fixed Galactocentric radius remain limited, largely because disentangling intrinsic chemical changes from those driven by radial migration is difficult.

Berkeley~32 is one of the oldest and most metal-poor clusters in the outer-to-intermediate disk, and a well-established benchmark of old disk populations \citep{Friel1995, Donati2015, Tang2017}. \citet{Sariya2021} provided a kinematic and structural baseline using Gaia DR2, identifying 563 probable members, determining an age of $4.90 \pm 0.22$~Gyr at a heliocentric distance of $3.76 \pm 0.17$~kpc, and confirming that the cluster is dynamically relaxed. However, that study was limited to Gaia DR2 astrometry and photometry, and did not incorporate spectroscopic abundances or a chemo-dynamical analysis linking orbital history to chemical enrichment. More recently, \citet{Donada2026} identified Berkeley~32 as a strong candidate for inward radial migration in the outer disk ($R_{\rm GC} > 9$~kpc), standing out as a clear deviation from the radial metallicity gradient, which motivates a comprehensive chemo-dynamical reanalysis.

This study presents a detailed reanalysis of Berkeley~32 by combining high-precision Gaia astrometry, homogeneous spectroscopic abundances, and orbital dynamics within a unified chemo-dynamical framework. While previous studies have examined its structural and chemical properties, a self-consistent analysis linking chemical clocks, orbital evolution, and birth-radius reconstruction remains limited. By combining these diagnostics, we provide empirical constraints on the chemical enrichment history of the Galactic disk at intermediate radii and the role of radial migration in shaping present-day abundance patterns. Berkeley~32 thus serves as a valuable benchmark for future studies of Galactic chemical evolution using combined chemical and dynamical diagnostics.

Section~\ref{sec:data} describes the Gaia DR3 and Gaia-ESO Survey Data Release 5.1 datasets. Section~\ref{sec:membership} presents the GMM-based membership analysis. Section~\ref{sec:rdp} derives the structural parameters of the cluster. Section~\ref{sec:fundamental} presents the isochrone-fitting results and the chemical abundance analysis. Section~\ref{sec:orbit} describes the orbital integration and birth-radius reconstruction. Section~\ref{sec:relax} investigates the dynamical state and internal dynamical evolution of the cluster. Section~\ref{sec:discussion} discusses the results in the context of Galactic chemical evolution, and Section~\ref{sec:conclusion} summarises the main conclusions.

\section{Data}\label{sec:data}

The \textit{Gaia} mission \citep{GaiaCollaboration2016} has revolutionized studies of the Milky Way by providing precise astrometric and photometric measurements. Gaia DR3 \citep{collaboration2023gaia} contains data for approximately 1.46 billion sources, including sky positions ($\alpha$, $\delta$), trigonometric parallaxes ($\varpi$), proper motions ($\mu_\alpha \cos\delta$, $\mu_\delta$), and $G$, $G_{\rm BP}$, and $G_{\rm RP}$ photometry \citep{Riello2021}. We queried the Gaia DR3 archive for all sources within a circular region of radius $r=18$~arcmin centered on Berkeley~32. Full details of the query and quality filters are provided in Appendix~\ref{app:data_quality}.

For the chemical abundance analysis, we used the Gaia-ESO Survey (GES) data release 5.1 \citep{Hourihane2023}, corresponding to the survey's sixth internal data release (iDR6). GES collected high-resolution spectra for more than $10^5$ stars using the VLT/FLAMES multi-object spectrograph \citep{Pasquini2002} between December 2011 and January 2018. Spectra were obtained with both the UVES ($R \approx 47,000$; setups U520 and U580, covering approximately 4140--6840~\AA) and GIRAFFE ($R \approx 17,000$--$31,000$; multiple HR setups) arms. Stellar parameters and elemental abundances were derived homogeneously across all working groups and placed on a common scale through the WG15 homogenization procedure described in \citet{Hourihane2023}.To ensure reliable abundance measurements, we retained only spectra with a signal-to-noise ratio of $\mathrm{SNR} \geq 30$. We further excluded sources flagged as \texttt{SRP}, indicating problems during the spectral reduction process, and \texttt{NIA}, corresponding to sources with too few spectral lines for reliable abundance determinations\footnote{\url{https://www.eso.org/rm/api/v1/public/releaseDescriptions/191}.}. In addition, for the 138 stars in our final sample, we verified that the membership probabilities from \citet{Jackson2022} lie, derived from Gaia EDR3 \citep{GaiaCollaboration2021} astrometry, in the range 0.82--1.00. Member stars identified in our Gaia DR3 analysis were cross-matched against the GES catalog using the VizieR XMatch service \citep{Boch2012} with a matching radius of $1\arcsec$.

An initial kinematic pre-selection retained stars with trigonometric parallaxes and proper motions close to the mean cluster values reported by \citet{Hunt2024}, thereby reducing field contamination before the formal membership analysis. These curated samples serve as input to the membership analysis (Section~\ref{sec:membership}) and are subsequently used for isochrone fitting, orbital integration, chemical abundance matching, and mass function determination.

\section{Membership Analysis}\label{sec:membership}

Identifying genuine cluster members against the Galactic field is essential for all subsequent analyses. Although reliable Gaia DR3 membership catalogues for Berkeley~32 are already available  \citep{Cantat-Gaudin2020, Hunt2024}, we derive an independent membership catalogue to provide a homogeneous and self-consistent sample for this study. Cluster membership is determined using a Gaussian Mixture Model \citep[GMM;][]{dempster1977maximum,mclachlan2000finite} applied to the Gaia DR3 astrometry of Berkeley 32. The GMM models the observed distribution of proper motions and trigonometric parallaxes as a combination of Gaussian components, one representing the cluster and one (or more) representing the field, and assigns a continuous membership probability to each star based on its location in astrometric space. This probabilistic framework is particularly effective in crowded Galactic fields where cluster and field populations exhibit substantial overlap in astrometric space. The method has been widely adopted in recent OC studies \citep{gao2018machine, agarwal2021ml, belwal2025unveiling, belwal2026time, bisht2026multiwavelength}.

\begin{figure*}
\centering
\includegraphics[width=0.95\linewidth]{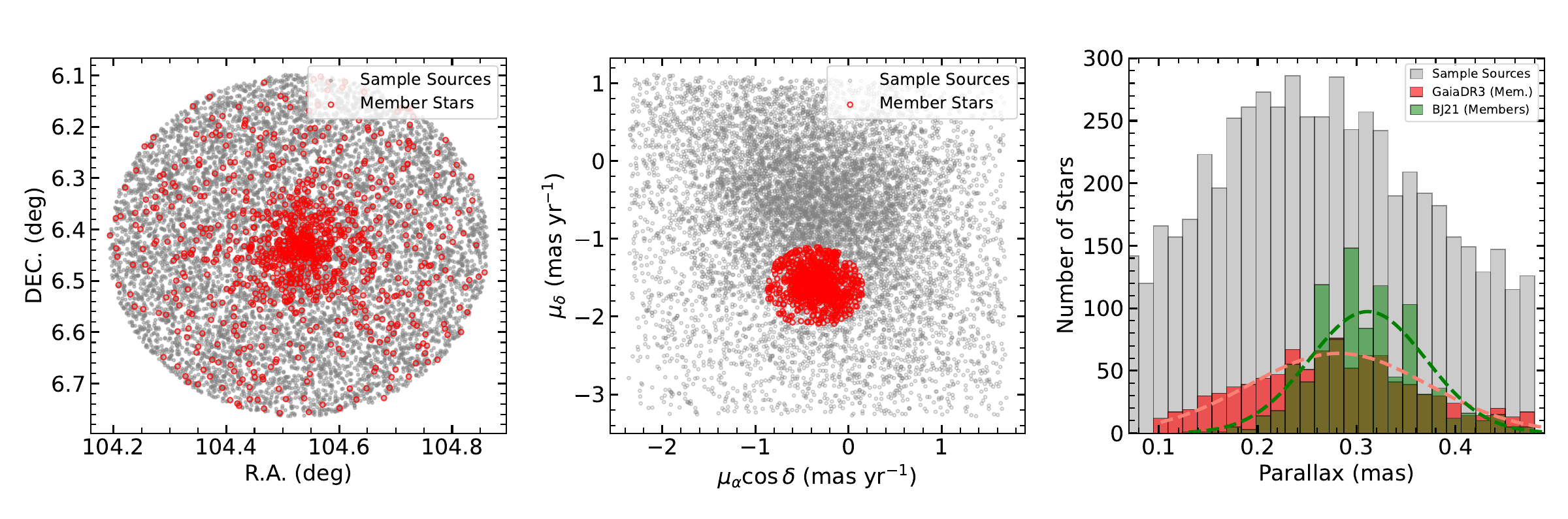}
\vspace{-0.5cm}
\caption{Spatial, kinematic, and astrometric distributions of Berkeley~32. The left panel displays the spatial distribution of sources in equatorial coordinates ($\alpha$, $\delta$), the middle panel presents the VPD highlighting the separation between the cluster bulk motion and the field, and the right panel shows the distribution of trigonometric parallaxes. In all panels, gray points and histograms represent the full sample of sources within the investigated field of view, while red circles and red filled bars denote the most probable cluster members identified using Gaia DR3 astrometry. Green-filled bars show the parallax distribution of the same members derived from \citet{BailerJones2021} photogeometric distances, with dashed pink and green curves representing Gaussian fits to the respective distributions.}
\label{fig:spatial_kinematic_dist}
\end{figure*}

Starting from Gaia DR3 sources within the search \textcolor{blue}{radius}, we retained only stars with reliable five-parameter astrometry and valid photometry in all three Gaia bands. Quality cuts include positive parallax ($\varpi > 0$), proper-motion uncertainties below a threshold $\epsilon_\mu$\textcolor{blue}{, and} $\mathrm{RUWE} \leq 1.4$ \citep{Lindegren2021}, which excludes stars with poor single-star astrometric solutions, likely unresolved binaries, or problematic measurements. To guide the GMM initialization and reduce contamination, a preliminary cluster locus in ($\mu_\alpha \cos\delta$, $\mu_\delta$, $\varpi$) space was estimated using the $k$-Nearest Neighbors algorithm \citep[kNN;][]{cover1967nearest}. Astrometric data were then standardized, and a two-component GMM was fitted via Expectation-Maximization. Each star received a membership probability $P$, and probable members were selected as those with $P \geq P_{\min}$, balancing completeness and field contamination for Berkeley~32.

Applying this procedure, we identify $N_{\rm mem} = 973$ probable members for Berkeley 32 with $P \geq 0.7$. The mean proper motions of the selected stars are consistent with literature values \citep{Cantat-Gaudin2020, Hunt2024}, and cross-matching with the \citet{Hunt2024} catalog confirms a large overlap, supporting the robustness of our selection. The sky distribution, proper-motion vector-point diagram (VPD), and trigonometric parallax histogram (Figure~\ref{fig:spatial_kinematic_dist}) show that the chosen members form compact overdensities both spatially and in kinematic space, with a narrow parallax peak well separated from the field distribution, confirming the GMM yields clean and internally consistent membership lists.

The center of Berkeley~32 was determined by identifying the region of highest stellar density on the sky. We used the equatorial coordinates ($\alpha,\delta$) of stars extracted from the Gaia DR3 catalogue. One-dimensional histograms were constructed separately for the $\alpha$ and $\delta$ distributions, and Gaussian functions were fitted to determine the peak positions. The maxima of these Gaussian fits were adopted as the refined equatorial coordinates of the cluster center, corresponding to the location of the highest projected stellar density. The final equatorial coordinates are $\alpha = 06^{\text{h}}58^{\text{m}}08.94^{\text{s}}$ $\delta = +06^\circ26^\prime00^{\prime\prime}\!.91$ corresponding to Galactic coordinates $l = 207^\circ\!.9550$ and $b = +4^\circ\!.4061$ for Berkeley~32. The probability threshold $p_{\rm min}$ was optimized to balance sample completeness and field contamination. We verified that moderate variations in $p_{\rm min}$ do not significantly affect the derived cluster parameters, ensuring the robustness of our results.

The mean astrometric parameters of Berkeley~32 were derived from the Gaia DR3 data of high-probability member stars. The mean proper-motion components were determined as $\langle\mu_{\alpha}\cos\delta, \mu_{\delta}\rangle = (-0.355 \pm 0.169, -1.589 \pm 0.154)$~mas~yr$^{-1}$, and a Gaussian fit to the parallax distribution yields a mean trigonometric parallax of $\varpi = 0.273 \pm 0.072$~mas, corresponding to a distance of $d_{\varpi} = 3.66 \pm 0.97$~kpc. The derived proper motions and parallax are in good agreement with previous determinations \citep{Cantat-Gaudin2020, Sariya2021, Hunt2024}; all astrometric parameters are summarised in Table~\ref{tab:summary}.

\begin{table*}
\centering
\footnotesize
\renewcommand{\arraystretch}{1.2}
\caption{Fundamental parameters derived for Berkeley~32 together with recent literature values.}
\label{tab:summary}
\begin{tabular}{llrrr}
\hline\hline
Parameter & Symbol & This study & \citet{Sariya2021} & \citet{Hunt2024} \\
\hline
\multicolumn{5}{c}{\textit{Astrometric Parameters}} \\
\hline
Right ascension                 & $\alpha$ (hh:mm:ss)  & 06:58:08.94  & 06:58:07.20 & 06:58:07.45 \\
Declination                     & $\delta$ (dd:mm:ss)  & +06:26:00.91 & +06:25:58.80 & +06:25:59.75 \\
Galactic longitude              & $l$ (degree)       & 207.9550                       & 207.9550 & 207.9547 \\
Galactic latitude               & $b$ (degree)       & +4.4061                        & +4.4090 & +4.4097 \\
Mean proper motion (RA)         & $\langle\mu_{\alpha}\cos\delta\rangle$ (mas yr$^{-1}$) & $-0.355 \pm  0.169$ & $-0.34 \pm 0.008$ & $-0.356 \pm 0.108$ \\
Mean proper motion (Dec)        & $\langle\mu_{\delta}\rangle$ (mas yr$^{-1}$)        & $-1.589 \pm 0.154$ & $-1.60 \pm 0.006$ & $-1.590 \pm 0.094$ \\
Mean trigonometric parallax     & $\langle\varpi\rangle$ (mas)                & $0.273 \pm 0.072$    & $0.280 \pm 0.004$ & $0.275 \pm 0.067$ \\
Mean parallax distance          & $\langle d_{\varpi}\rangle$ (kpc)       & $3.66 \pm 0.97$    & $3.57\pm 0.05$ & $3.148^{+0.022}_{-0.022}$ \\
\hline
\multicolumn{5}{c}{\textit{Structural Parameters}} \\
\hline
Central surface density         & $\rho_0$ (stars arcmin$^{-2}$)    & $51.88^{+5.16}_{-4.54}$ & $11.47$ & $--$ \\
Background density              & $\rho_{\rm bg}$ (stars arcmin$^{-2}$) & $0.39^{+0.05}_{-0.05}$ & $0.08$ & $--$ \\
Core radius                     & $r_{\rm c}$ (arcmin)       & $1.30^{+0.13}_{-0.13}$   & $1.71$ & $3.21$ \\
Tidal radius                    & $r_{\rm t}$ (arcmin)       & $9.06^{+0.24}_{-0.24}$  & $9.40$ & $15.66$ \\
Half-mass radius                & $R_{\rm h}$ (pc)       & $1.95$                                   & $--$ & $3.45$ \\
Concentration parameter         & $C$         & $0.84$                                      & $0.74$ & $--$ \\
\hline
\multicolumn{5}{c}{\textit{Fundamental Parameters}} \\
\hline
Photogeometric distance         & $\langle d_{\rm pgeo}\rangle$ (kpc) & $3.325 \pm 0.634$  & $3.76 \pm 0.17$ & $3.148^{+0.022}_{-0.022}$ \\
True distance modulus           & $(m-M)_0$ (mag)                     & $12.970\pm 0.033$  & $12.876$ & $12.490$ \\
Apparent distance modulus       & $(m-M)$ (mag)                       & $13.196 \pm 0.081$ & $13.10 \pm 0.02$ & $12.540^{+0.158}_{-0.203}$ \\
Mean extinction                 & $\langle A_{\rm V} \rangle$ (mag)   & $0.383 \pm 0.125$  & $0.284$ & $0.163^{+0.065}_{-0.143}$ \\
Color excess ($UBV$)            & $E(B-V)$ (mag)                      & $0.124 \pm 0.040$  & $0.092$ & $0.053^{+0.021}_{-0.046}$ \\
Color excess (Gaia)             & $E(G_{\rm BP}-G_{\rm RP})$ (mag)    & $0.175 \pm 0.056$  & $0.13$ & $0.075^{+0.030}_{-0.066}$ \\
$G$-band extinction             & $A_{\rm G}$ (mag)                   & $0.226 \pm 0.074$  & $0.224$ & $0.144^{+0.057}_{-0.126}$ \\
Iron abundance                  & $[\rm{Fe/H}]$ (dex)                 & $-0.39 \pm0.02$    & $-0.38$ & $--$ \\
Metallicity                     & $Z$                                 & $0.0064$           & $0.0060$ & $--$ \\
Age (isochrone)                 & $t_{\rm iso}$ (Gyr)                 & $4.90 \pm 0.50$             & $4.90 \pm 0.22$ & $3.25^{+1.59}_{-0.83}$ \\
Age ([Y/Mg] clock)              & $t_{\rm [Y/Mg]}$ (Gyr)              & $4.73 \pm 2.39$    & $--$ & $--$ \\
\hline
\multicolumn{5}{c}{\textit{Galactic Orbital Parameters}} \\
\hline
Mean radial velocity            & $V_{\rm rad,OC}$ (km s$^{-1}$)        & $106.26 \pm 0.03$  & $106.40 \pm 0.47$ & $103.04 \pm 2.51$ \\
Perigalactic radius             & $R_{\rm peri}$ (kpc)              & $6.911 \pm 0.086$          & $12.18$ & $--$ \\
Apogalactic radius              & $R_{\rm apo}$ (kpc)               & $11.971 \pm 0.059$         & $12.28$ & $--$ \\
Mean orbital radius             & $R_{\rm mean}$ (kpc)              & $9.441 \pm 0.073$          & $12.23$ & $--$ \\
Orbital eccentricity            & $e$                               & $0.268 \pm 0.004$          & $0.00$  & $--$ \\
Maximum vertical height         & $Z_{\rm max}$ (kpc)               & $0.311 \pm 0.008$          & $0.32$  & $--$ \\
Present Galactocentric radius   & $R_{\rm GC}$ (kpc)                & $11.107 \pm 0.049$         & $11.74$ & $--$ \\
Guiding radius                  & $R_{\rm g}$ (kpc)                 & $8.821 \pm 0.085$          & $--$    & $--$ \\
Chemical birth radius           & $R_{\rm b}$ (kpc)                 & $9.822$                    & $--$    & $--$ \\
Orbital period                  & $T_{\rm p}$ (Myr)                 & $272 \pm 2$                & $376$   & $--$ \\
\hline
\multicolumn{5}{c}{\textit{Dynamical Parameters}} \\
\hline
Relaxation time                 & $T_{\rm relax}$ (Myr)             & $30$                       & $109 \pm 7$ & $--$ \\
Dynamical age parameter         & $\tau$                            & $164$                      & $--$      & $--$ \\
Total cluster mass              & $M_{\rm C}$ ($M_\odot$)           & $1125$                     & $--$      & $3540 \pm 348$ \\
Mean stellar mass               & $\langle M_C \rangle$ ($M_\odot$) & $1.00$                     & $0.92$    & $--$ \\
\hline
\end{tabular}

\end{table*}

\section{Structural Parameters Determination}
\label{sec:rdp}

To characterize the cluster's internal structure, we analyzed the radial stellar density distributions of high-probability cluster members identified through our Gaia astrometric analysis. The radial density profile (RDP) was constructed using an equal-area annular binning scheme, ensuring uniform sampling of the stellar density field and comparable Poisson uncertainties across bins, with each annulus containing at least 20 member stars ($N_i \geq 20$).

The surface density $\rho(r_i)$ at projected distance $r_i$ from the cluster center was obtained as $\rho(r_i) = N_i / A_i$, where $N_i$ is the star count and $A_i$ the annular area. We fitted the observed profiles with the empirical surface-density model of \citet{King62}:
\begin{equation}
\rho(r) = \rho_0 \left[ \frac{1}{\sqrt{1 + (r/r_{\rm c})^2}} - \frac{1}{\sqrt{1 + (r_{\rm t}/r_{\rm c})^2}} \right]^2 + \rho_{\rm bg},
\label{eq:king_model}
\end{equation}
where $\rho_0$ is the central density, $r_{\rm c}$ is the core radius (where $\rho = 0.5\,\rho_0$), $r_{\rm t}$ is the tidal radius, and $\rho_{\rm bg}$ is the residual background density. Although $\rho_{\rm bg} \approx 0$ is expected for a member-only sample, it was retained as a free parameter to account for residual contamination. Best-fit parameters were estimated by minimizing the negative log-likelihood:
\begin{equation}
\ln \mathcal{L} = -\frac{1}{2} \sum_i \left( \frac{\rho_i - \rho_{i,{\rm model}}}{\sigma_{\rho_i}} \right)^2,
\label{eq:log_likelihood}
\end{equation}
where $\sigma_{\rho_i}$ is the Poisson uncertainty in each annulus.

\begin{figure}
    \centering
    \includegraphics[width=1\linewidth]{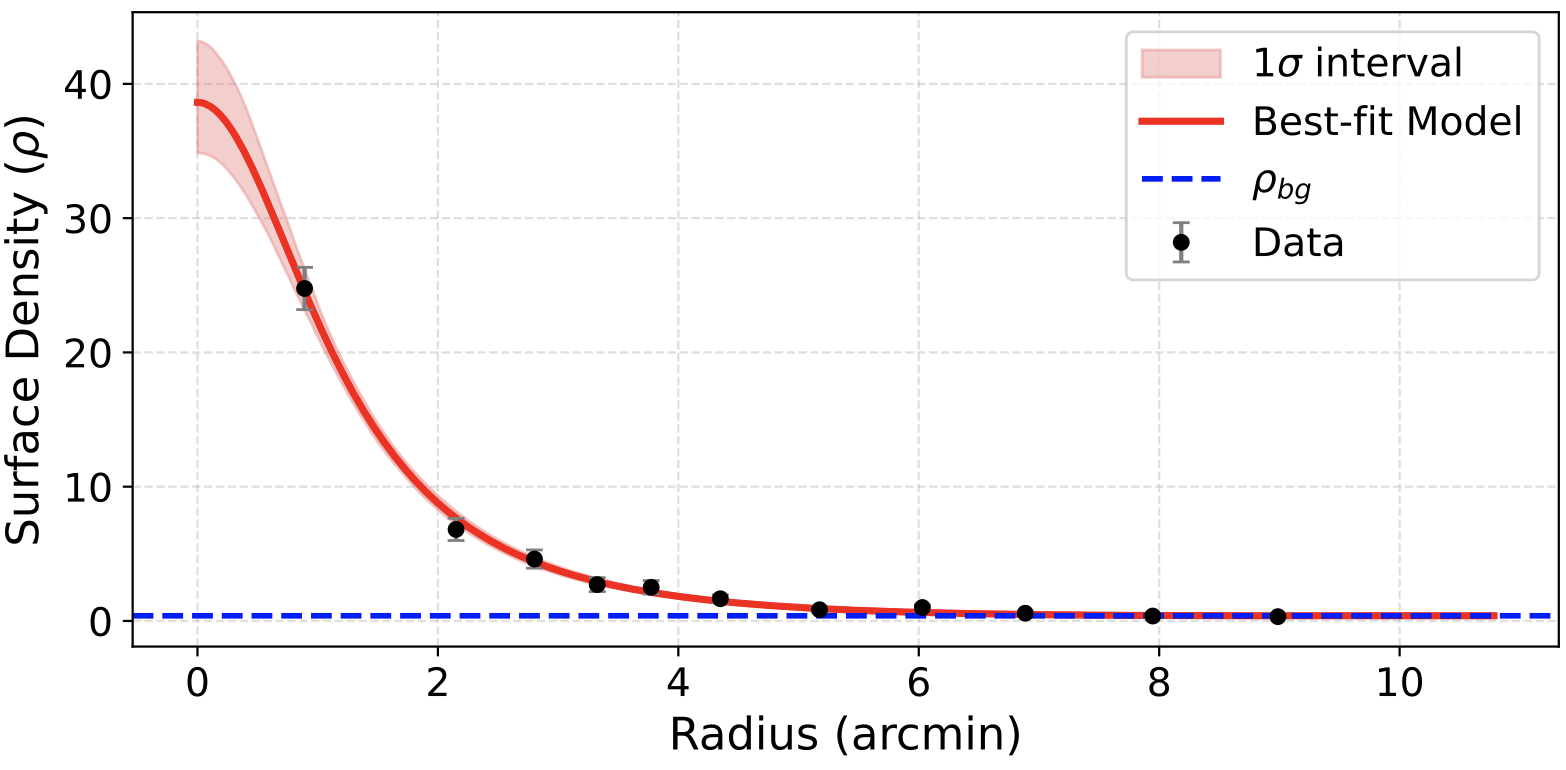}\\
    \includegraphics[width=1\linewidth]{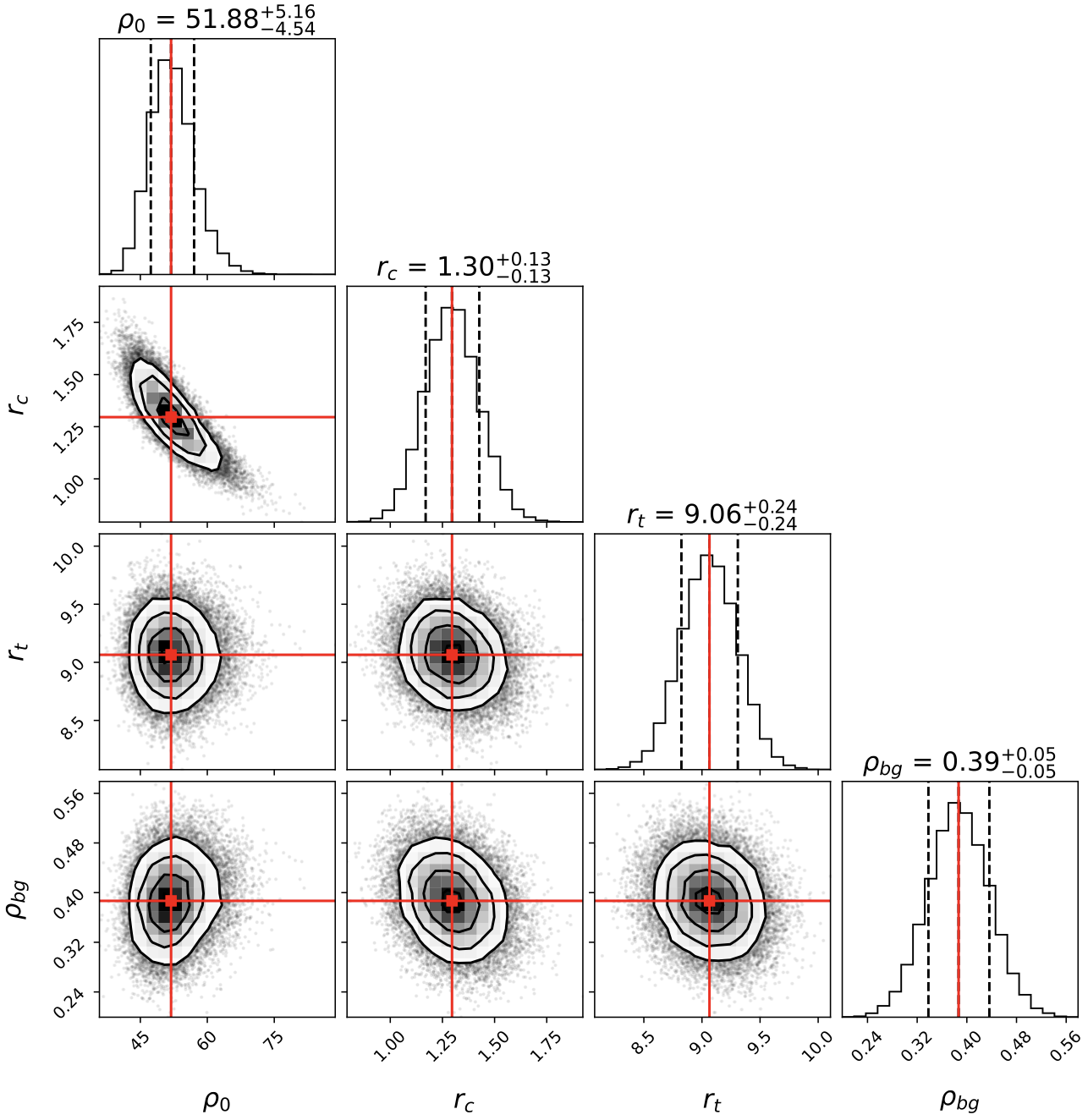}
\caption{RDP and MCMC posterior distributions for Berkeley~32. \textit{Upper panel:} Observed surface density ($\rho$) as a function of radial distance from the cluster center (black points with Poisson error bars). The solid red curve is the best-fitting \citet{King62} model, with the shaded region indicating the $1\sigma$ confidence interval. The horizontal blue dashed line marks the background density level ($\rho_{\rm bg}$). \textit{Lower panels:} Corner plot of the posterior distributions for the fitted King model parameters ($\rho_0$, $r_{\rm c}$, $r_{\rm t}$, $\rho_{\rm bg}$), with median values (solid red lines) and $1\sigma$ boundaries (dashed lines).}
    \label{fig:RDP_fits}
\end{figure}

Parameter optimization was performed using the MCMC ensemble sampler \texttt{emcee} \citep{emcee} with 100 walkers, 5000 steps, and a 500-step burn-in, adopting broad uniform priors on all parameters. Chain convergence was confirmed via the \citet{gelman1992} $\hat{R}$ diagnostic ($\hat{R} < 1.1$). The resulting best-fit models and $1\sigma$ confidence intervals are shown in Figure~\ref{fig:RDP_fits}, and the structural parameters are summarised in Table~\ref{tab:summary}.

Following \citet{King62}, the concentration parameter $C = \log(r_{\rm t}/r_{\rm c})$ indicates a cluster's physical compactness and dynamical state. For Berkeley~32, we obtain $C \approx 0.84$ ($r_{\rm t}/r_{\rm c} \approx 7$), characteristic of a moderately concentrated, dynamically relaxed OC.

\section{Color-Magnitude Diagram and Derived Cluster Parameters}\label{sec:fundamental}
Reliable determination of the fundamental parameters of Galactic OCs is essential for understanding the formation and evolution of stellar clusters, as well as the structure and chemical evolution of the Galactic disk \citep{Friel1995, Netopil2016}. However, deriving the fundamental physical parameters of OCs, such as age, distance, reddening, and metallicity, remains a complex task. In particular, clusters located at low Galactic latitudes are significantly affected by interstellar extinction and often exhibit differential reddening across the cluster field \citep{Burki1975, Carraro2017}. In addition, Galactic OCs display a broad metallicity distribution, with variations among clusters reaching up to $\sim$0.25 dex \citep{Cinar2024, Tasedemir2026}.

In isochrone fitting, reddening, distance modulus, metallicity, and age are strongly correlated, meaning that different combinations of these parameters can produce very similar CMD morphologies \citep[e.g.,][]{vonHippel2006, Bilir2006a, Bilir2010, Andreuzzi2011, Akbulut2021}. This well-known degeneracy has been discussed extensively in the literature \citep{Yontan2015, Bostanci2015, Ak2016, Bostanci2018, Yontan2019, Cakmak2024} and can lead to non-unique or physically inconsistent solutions when additional observational constraints are not taken into account. Therefore, it is necessary to constrain reddening, distance, and metallicity through independent methods before performing the final CMD-based age estimation.

To minimise the impact of the well-known degeneracy between reddening, distance, metallicity, and age in isochrone fitting, we independently constrained these parameters prior to the final CMD analysis. Line-of-sight $V$-band extinction values were derived on a star-by-star basis using the three-dimensional dust maps of \citet{Green2019}, stellar distances were adopted from the Bayesian photogeometric catalogue of \citet{BailerJones2021}, and the cluster metallicity was constrained using high-resolution GES spectroscopic data. By fixing these three quantities independently, the subsequent isochrone fitting reduces to a single optimization parameter: the cluster age.

\subsection{Photogeometric Distance}\label{sec:photogeometric_distance}

Since Berkeley~32 is located at approximately 3~kpc, directly inverting raw parallax values for individual cluster members can yield biased and physically inconsistent distance estimates \citep{Plevne2020}. We therefore adopt the Bayesian photogeometric distance estimates of \citet{BailerJones2021}, which incorporate both parallax measurements and photometric information within a probabilistic framework that accounts for a direction-dependent prior based on a three-dimensional model of the Milky Way. This approach provides more stable and reliable distance estimates, particularly for distant sources with low-precision parallaxes. The catalogue of \citet{BailerJones2021} provides Bayesian distance estimates for approximately 1.47 billion \textit{Gaia} EDR3 sources, offering two types of estimates for each star: the purely geometric distance ($d_{\rm geo}$), which relies predominantly on trigonometric parallax, and the photogeometric distance ($d_{\rm pgeo}$), which additionally incorporates stellar magnitudes and color indices. Since $d_{\rm pgeo}$ values are generally more tightly constrained and exhibit smaller uncertainties, particularly for stars with low-precision parallaxes, we adopt $d_{\rm pgeo}$ throughout this study.

Because the membership catalogue does not include Gaia EDR3 source identifiers, the cluster members were matched with the catalogue of \citet{BailerJones2021} using their sky coordinates. For this purpose, a cone search around the cluster field was performed using the VizieR service, and the sources were cross-matched within a 1$^{\prime\prime}$ radius, consistent with Gaia's astrometric precision. Only confirmed cluster members were retained in the final sample. The parallax equivalents of the photogeometric distances of the member stars taken from the \citet{BailerJones2021} catalogue are shown in the right-hand panel of Figure~\ref{fig:spatial_kinematic_dist}. The median $d_{\rm pgeo}$ of the cluster was derived directly from the individual $d_{\rm pgeo}$ values of \citet{BailerJones2021} as $\langle d_{\rm pgeo} \rangle = 3325 \pm 634$~pc, consistent with Berkeley~32 being a distant, old OC located well beyond the solar neighbourhood.

\subsection{Interstellar Extinction}
Interstellar reddening is strongly coupled with distance and age in isochrone fitting, making reliable extinction estimates essential for reducing parameter degeneracies. We therefore adopt a distance-dependent three-dimensional dust map to estimate the line-of-sight extinction toward individual cluster members, providing a more physically consistent characterization than assuming a single mean extinction for the entire cluster. In this approach, the interstellar reddening of cluster members was estimated using the three-dimensional dust map of \citet{Green2019}, {\sc Bayestar19}, which adopts the reddening calibration of \citet{Schlafly2011}. This map is particularly well-suited for OC studies, as it provides distance-dependent extinction estimates based on Pan-STARRS 1 and 2MASS photometry, and is broadly consistent with \textit{Gaia} astrometric constraints. For each member star, the line-of-sight reddening $E(B-V)$ was interpolated at its sky position, and the photogeometric distance ($d_{\rm pgeo}$), and the median value was adopted to mitigate the effects of small-scale spatial variations in the dust distribution. The original {\sc Bayestar19} reddening values, expressed in terms of $E(G_{\rm BP}-G_{\rm RP})$, were converted to $E(B-V)$ using the transformation $E(G_{\rm BP}-G_{\rm RP}) = 1.41 \times E(B-V)$ \citep{Canbay2023}, and the corresponding visual extinction was then derived as $A_{\rm V} = 3.1 \times E(B-V)$ \citep{Cardelli1989}.

\begin{figure}
    \centering
    \includegraphics[width=0.92\linewidth]{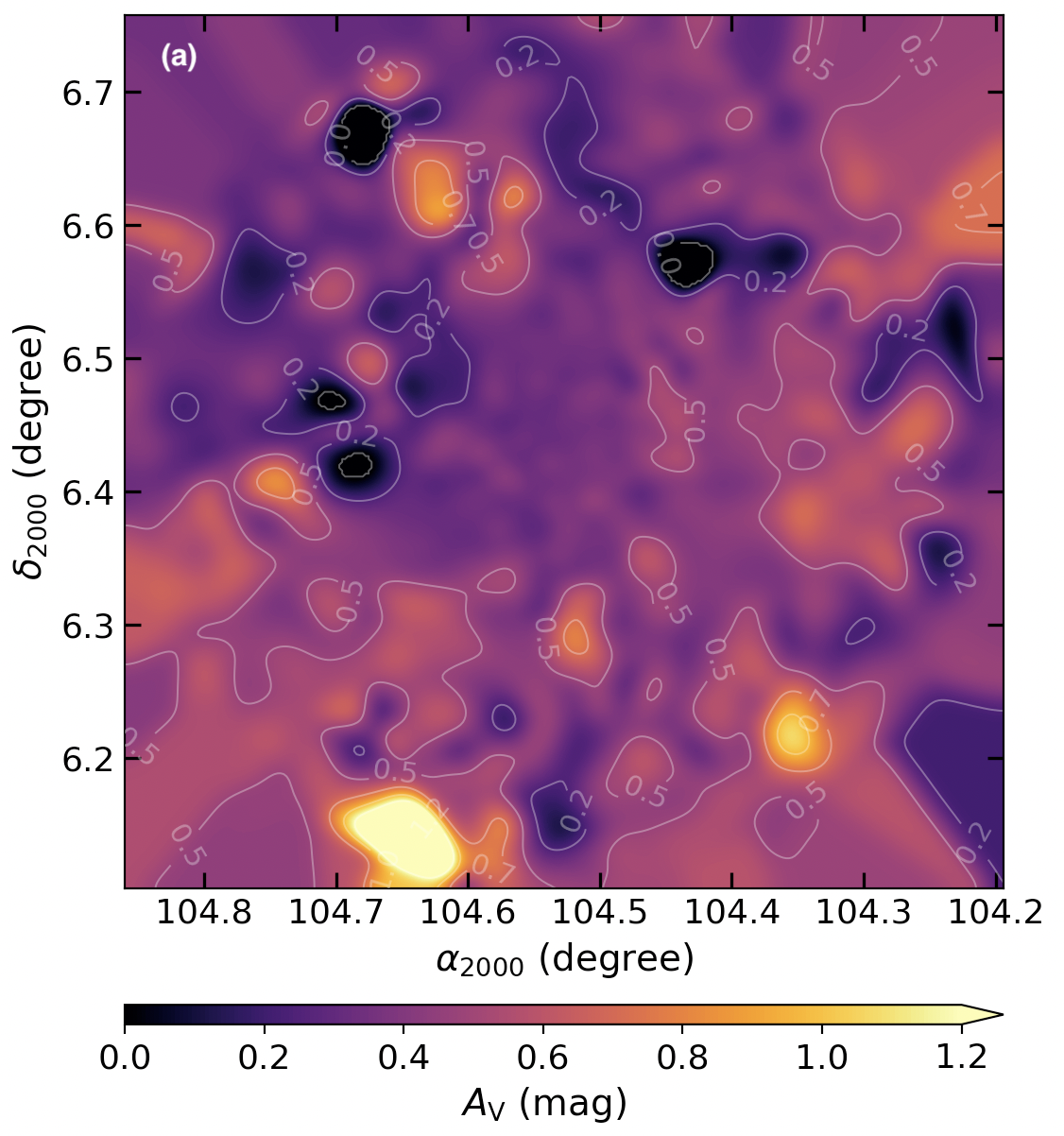}\\
    \includegraphics[width=0.87\linewidth]{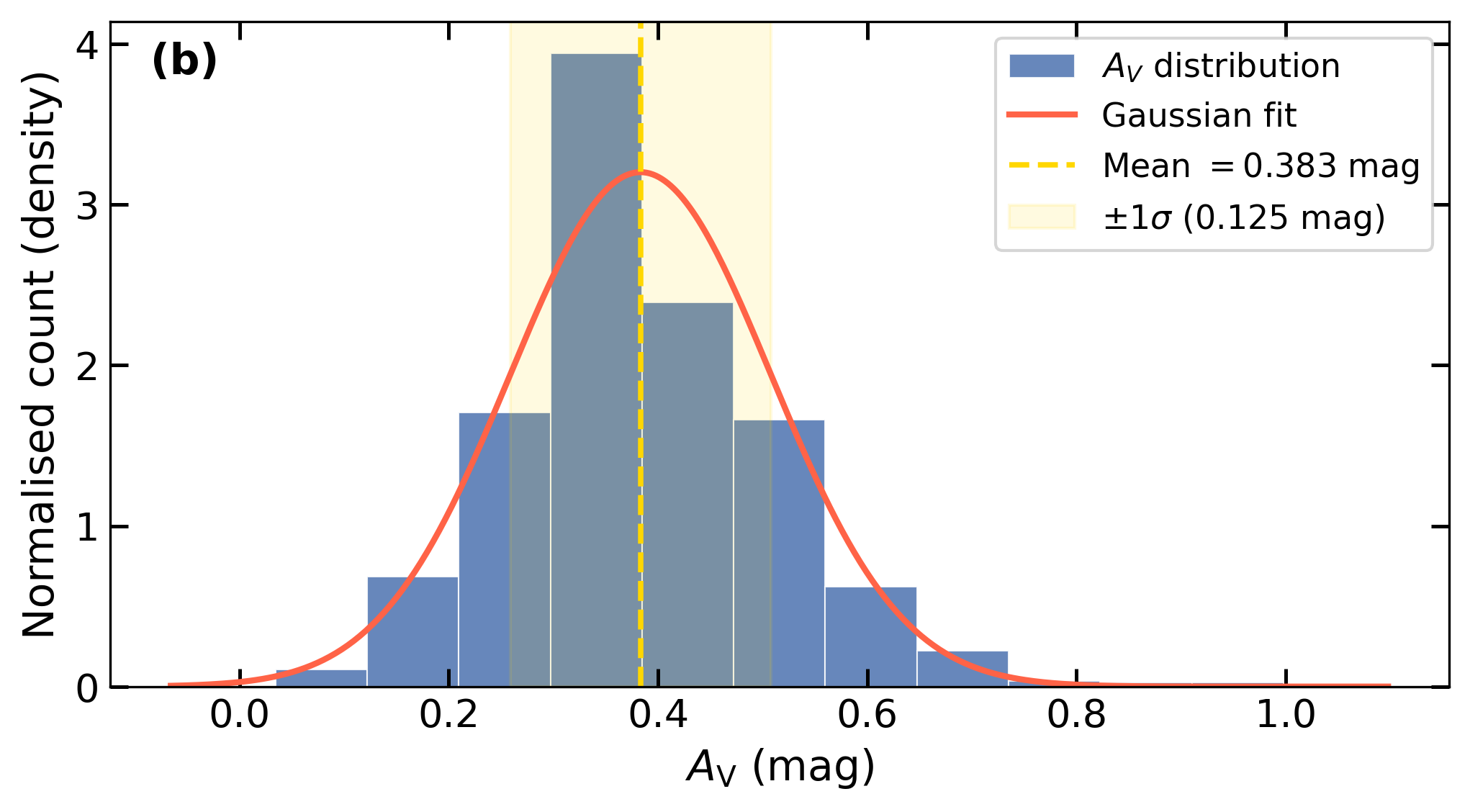}
    \caption{Extinction map and distribution of the cluster member stars. (a) Spatial variation of $A_{\rm V}$ (in magnitudes) derived from the {\sc Bayestar19} dust model \citep{Green2019}. Contours and color scale indicate the extinction level across the cluster field. (b) Distribution of individual $A_{\rm V}$ values for member stars. The red curve shows the Gaussian fit to the distribution, while the dashed vertical line marks the mean value. The shaded region indicates the $\pm1\sigma$ interval.}
    \label{fig:av_map}
\end{figure}

The spatial distribution of extinction across the cluster field is shown in Figure~\ref{fig:av_map}. The extinction map reveals spatially non-uniform reddening, with localized regions reaching $A_{\rm V}\sim1.2$ mag, while most cluster members lie in areas of lower extinction. The distribution of individual $A_{\rm V}$ values (Figure~\ref{fig:av_map}b) is well described by a Gaussian profile, suggesting smoothly varying extinction across the field. A Gaussian fit yields a mean extinction of $\langle A_{\rm V} \rangle = 0.383 \pm 0.125$ mag, corresponding to $E(B-V) = 0.124 \pm 0.040$ mag, $E(G_{\rm BP}-G_{\rm RP}) = 0.175 \pm 0.056$ mag, and $A_{\rm G} = 0.226 \pm 0.074$ mag.

\subsection{Metallicity}
\begin{figure*}
\centering
\includegraphics[width=0.4\linewidth]{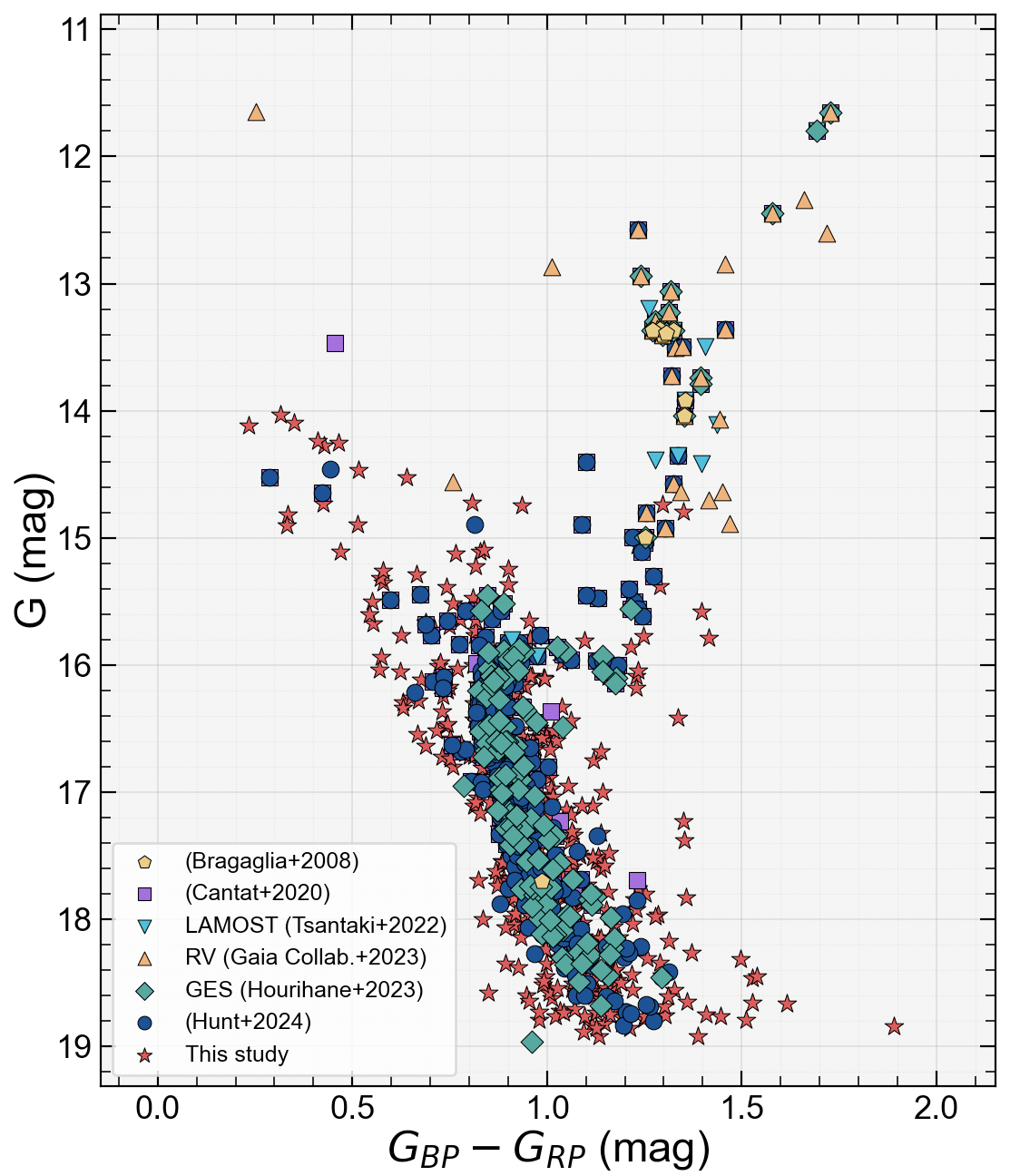}
\includegraphics[width=0.52\linewidth]{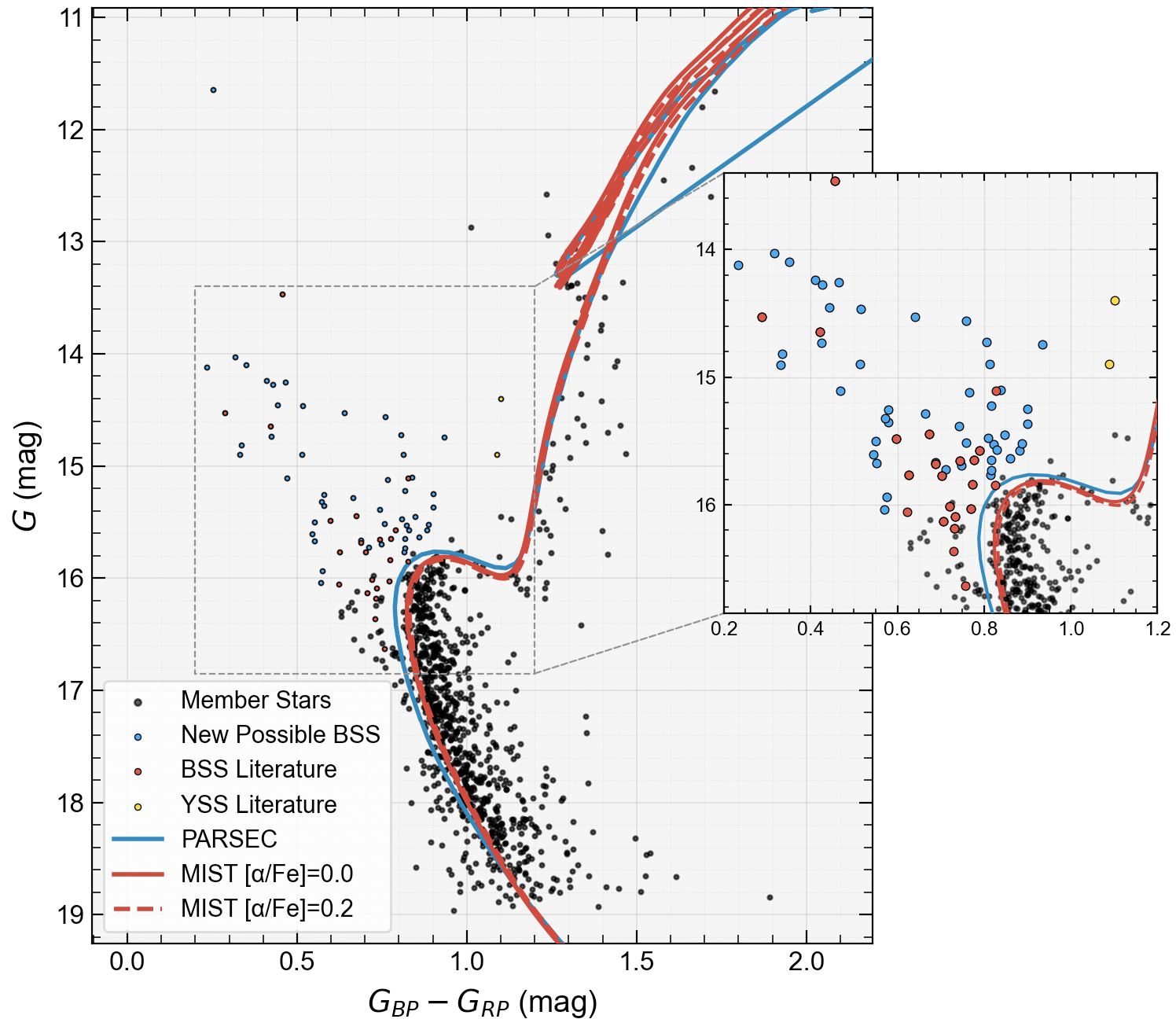}
\caption{CMDs of the high-probability cluster members. Members are plotted and color-coded according to cross-matched literature catalogs, as indicated in the legend. The red stars represent the members identified in this study (left panel).  Member stars are displayed as black points (right panel). Red and yellow symbols indicate literature BSS and YSS stars, while blue symbols show the photometric BSS candidates identified in this study. The solid curves represent the best-fitting theoretical PARSEC (blue) and MIST (red) isochrones. Their corresponding upper and lower age uncertainties are indicated by the dashed lines, demonstrating the consistency between the observed stellar sequences and the derived fundamental parameters for both models.}
\label{fig:CMDs}
\end{figure*}

Spectroscopic metallicities and $\alpha$-element abundances were adopted from GES DR5.1 \citep{Hourihane2023}, providing robust constraints for isochrone selection and chemical analysis \citep{Carrera2011}. Of the high-probability cluster members, 138 were matched with the GES DR5.1 catalog, yielding 431 spectra. Applying a quality cut of $\mathrm{SNR} \geq 30$ resulted in a final sample of 73 stars with 254 spectra, spanning $G = 11.5$--17.5 mag. These stars are primarily located along the main sequence (MS) and evolved main-sequence turn-off (MSTO) regions of the CMD (Figure~\ref{fig:CMDs}).

\begin{figure}
    \centering
    \includegraphics[width=0.95\linewidth]{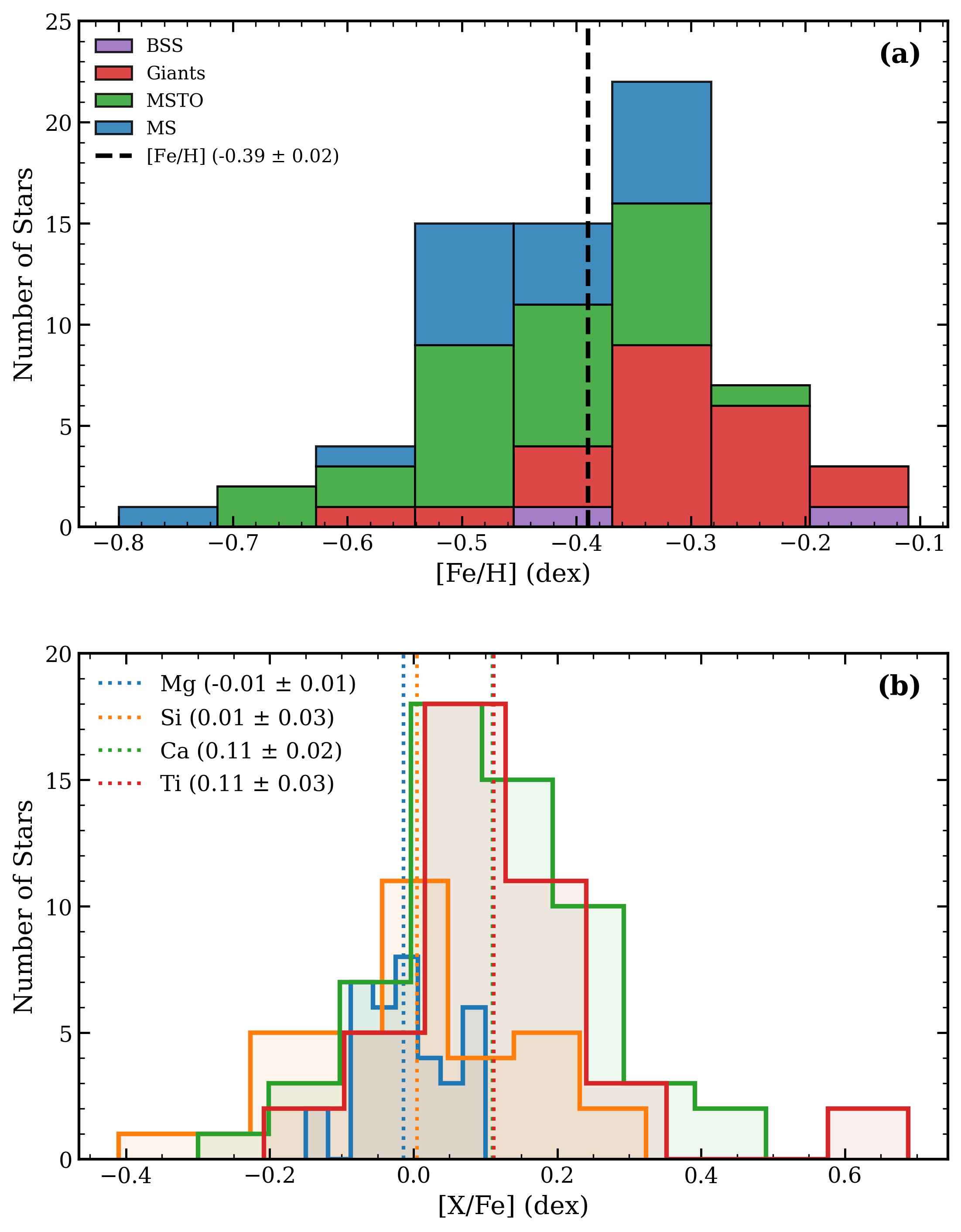}
    \caption{Element abundance distributions of Berkeley~32 member stars with SNR$\geq30$ derived from the GES data. Panel (a) presents the [Fe/H] distribution, where different colors within each histogram bin denote the number of stars belonging to different evolutionary stages. Panel (b) shows the abundance distributions of the $\alpha$ elements Mg, Si, Ca, and Ti. The vertical dashed lines represent the median values of the distributions, while the quoted uncertainties correspond to the standard error of the median.}
    \label{fig:element_histo}
\end{figure}

Using the [Fe/H] and [$\alpha$/Fe] abundances of 73 GES stars with reliably measured spectroscopic parameters and confirmed cluster membership, the chemical abundance properties of Berkeley 32 were investigated. For this purpose, histograms of the [Fe/H] and [$\alpha$/Fe] distributions were constructed. In the [Fe/H] histogram presented in Figure~\ref{fig:element_histo}a, stars are color-coded according to their evolutionary classes. As seen from the Figure~\ref{fig:element_histo}a, the majority of stars with measured iron abundances are located in the MSTO and MS evolutionary stages. The individual iron abundances are distributed approximately within the range of $-0.8 \leq \mathrm{[Fe/H]~(dex)} \leq -0.1$. The median iron abundance of Berkeley 32 was derived as $\langle \rm [Fe/H] \rangle = -0.39$ dex, while the standard deviation of the distribution was found to be 0.02 dex.

However, spectral line measurements for all $\alpha$ elements are not uniformly available in the literature. Therefore, only Mg, Si, Ca, and Ti were considered, with \citet{Hourihane2023} reporting measured abundances for 36, 34, 59, and 41 member stars, respectively. The analysis yields median abundance ratios of [Mg/Fe] = $-0.01 \pm 0.01$, [Si/Fe] = $0.01 \pm 0.03$, [Ca/Fe] = $0.11 \pm 0.02$, and [Ti/Fe] = $0.11 \pm 0.03$ dex, giving a mean [$\alpha$/Fe] = $0.06 \pm 0.01$ dex (Figure~\ref{fig:element_histo}b). This near-solar $\alpha$-element pattern suggests that Berkeley~32 formed from interstellar material enriched by both core-collapse and Type~Ia supernovae, consistent with the abundance properties commonly observed in old Galactic OCs \citep[e.g.][]{Magrini2009, Donati2015}. The mild enhancement in Ca and Ti may reflect a limited contribution from $\alpha$-capture nucleosynthesis \citep{Mcwilliam1997, Tolstoy2009}, while the slightly sub-solar Mn abundance may indicate a weaker or delayed contribution from Type~Ia supernovae \citep{Kobayashi2006, Bensby2014, battistini2015}.

To ensure consistency with stellar evolution models, the spectroscopically derived iron abundance was converted into the metallicity \citep[e.g.,][]{Karaali2011,Yontan2021, Yontan2022, Yontan2023b,  Alzhrani2025, Elsanhoury2026}. Adopting $\langle {\rm [Fe/H]} \rangle = -0.39 \pm 0.02$~dex, which corresponds to the mean iron abundance derived for Berkeley 32 in this study, and using the calibrated relation \citep[e.g.,][]{Yontan2023a,Gokmen2023, Bilir2026}:
\begin{equation}
Z_{\rm x} = \frac{Z}{0.7515 - 2.78 \times Z}
\end{equation}
\begin{equation}
\text{[Fe/H]} = \log (Z_{\rm x}) - \log \left( \frac{Z_{\odot}}{1 - 0.248 - 2.78 \times Z_{\odot}} \right),
\end{equation}
with $Z_{\odot}=0.0152$, we obtain a heavy-element mass fraction of $Z=0.0064\pm0.0003$. This value is adopted in the subsequent isochrone fitting.

\subsection{Isochrone Fitting}

The CMD is a primary observational tool for deriving the fundamental parameters of star clusters through comparison with theoretical stellar evolution models. In OCs, well-defined evolutionary sequences traced by high-probability member stars provide strong constraints for isochrone fitting, enabling accurate estimates of age, distance modulus, reddening, and chemical composition \citep[e.g.,][]{Carrera2011, Randich2018}. The left panel of Figure~\ref{fig:CMDs} presents the Gaia-based $G$ versus $G_{\rm BP}-G_{\rm RP}$ CMD of Berkeley~32, constructed using the high-probability cluster members identified through our astrometric membership analysis. The CMD exhibits a well-defined cluster sequence extending from the MS through the MSTO region to the red giant branch (RGB) and red clump. To assess the reliability of our membership selection, we cross-matched our high-probability member sample with several independent literature catalogs, including \citet{Cantat-Gaudin2020}, \citet{Hunt2024}, the radial velocity catalog of \citet{GaiaCollaboration2023}, the Survey of Surveys \citep[SoS;][]{Tsantaki2022}, element abundances from \citet{Bragaglia2008}, and the high-resolution spectroscopic analyses provided by the GES \citep{Hourihane2023}. The stellar sequences defined by our astrometric selection show remarkable consistency with these independent samples across all magnitude ranges, confirming the robustness of our membership criteria.

Before isochrone fitting, the fundamental cluster parameters required for the CMD analysis were determined independently using different observational methods. The cluster metallicity was derived from spectroscopic abundance analysis, while the extinction and distance were obtained from photometric and astrometric data. For Berkeley~32, a heliocentric distance of $\langle d\rangle=3325\pm 634$~pc, a Gaia $G$-band extinction of $\langle A_{\rm G}\rangle=0.226\pm 0.074$~mag, and a metallicity of $\langle [{\rm Fe/H}]\rangle=-0.39\pm 0.02$~dex were adopted. Using these parameters as fixed inputs, the observed CMD morphology was compared with theoretical stellar-evolution models to infer the cluster age.

\begin{figure*}
    \centering
    \includegraphics[width=1\linewidth]{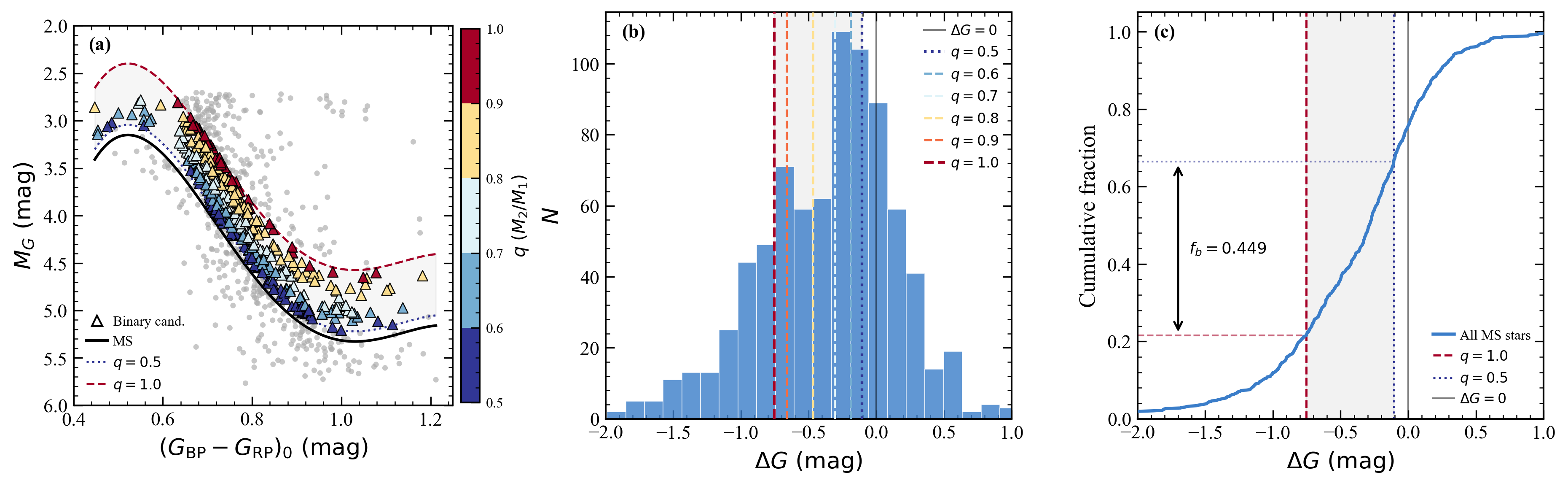}
    \vspace{-0.5cm}
    \caption{Binary fraction analysis of Berkeley~32. (a) CMD of the MS stars. The solid line shows the MS; the dashed and dotted lines mark the lower and upper limits of the binary selection region, respectively. Triangle symbols indicate binary candidates. (b) Distribution of $\Delta G$ residuals. The shaded region denotes the binary selection zone between the lower and upper limits. (c) Cumulative distribution of $\Delta G$. The binary fraction $f_b$ is indicated by the vertical arrow spanning the two selection limits.}
    \label{fig:berk32_binary}
\end{figure*}

We compared the observed CMD sequences with theoretical PARSEC CMD 3.9\footnote{\url{https://stev.oapd.inaf.it/cgi-bin/cmd}} \citep{Bressan2012} and MIST v2.5\footnote{\url{https://mist.science/interp_isos.html}} \citep{Choi2016, Dotter2016} isochrones. For the MIST models, we adopted an initial rotation rate of $v/v_{\rm crit}=0.4$ and $\alpha$-enhancement values of $[\alpha/{\rm Fe}]=0.0$ and $0.2$~dex. Both isochrone sets, evaluated at the same spectroscopic metallicity, provide consistent fits to the observed CMD morphology (Figure~\ref{fig:CMDs}), indicating that the derived parameters are not strongly dependent on the adopted stellar evolution model. The MIST turn-off lies at marginally redder colors than the PARSEC models, reflecting small systematic differences in the underlying stellar physics, including updated solar calibrations, rotation, and convective overshooting; similar model dependencies have been reported for younger clusters \citep{Bastian2025}. Varying the $\alpha$-element enhancement within the MIST set does not significantly alter the CMD morphology, confirming that the derived age is insensitive to the adopted $\alpha$-abundance over the explored range. The best-fitting isochrone solutions yield an age of $4.9 \pm 0.5$~Gyr for Berkeley~32, with both model grids simultaneously reproducing the MS, TO, and RGB within the quoted uncertainties. The age derived here is further compared with independent estimates in Section~\ref{sec:Ymg_age}.

\subsection{Binary Fraction and Stellar Content}\label{sec:stellar_content}

The binary fraction in OCs provides important constraints on their dynamical evolution, internal kinematics, and long-term survival \citep[e.g.][]{vonHippel2002, Sollima2007}. In particular, unresolved binary systems can significantly affect the morphology of the CMD, biasing the interpretation of stellar populations if not properly accounted for \citep{Milone2012}. In this context, we investigate the binary content of Berkeley~32 using a photometric approach following the methodology of \citet{Milone2012} and \citet{Donada2023}. 

For this purpose, MS stars were selected between the MSTO magnitude ($G_{\rm MSTO}=15.9$ mag) and 4 mag below it, within the color index interval $0.6 \leq (G_{\rm BP}-G_{\rm RP})~{\rm (mag)} \leq 1.4$. After applying these selection criteria, the final MS sample consists of 822 stars. The single-star sequence was defined by fitting a fourth-degree polynomial to the median MS locus in the CMD. To minimise the influence of differential reddening, the analysis was carried out in the extinction-corrected, absolute magnitude plane, using de-reddened color indices $(G_{\rm BP}-G_{\rm RP})_0$ and absolute magnitudes $M_{\rm G}$ derived from the cluster distance modulus and mean extinction values (see Table~\ref{tab:summary}). The brightness  offset between a binary system and a single star was computed using the relation:
\begin{equation}
    \Delta m = -2.5\log_{10}(1 + q^{\alpha}),
\end{equation}
where $\alpha = 3.5$ is the standard mass--luminosity exponent for MS stars, and $q = M_2/M_1$ is the mass ratio \citep{Eker2015, Eker2018, Eker2024}. 

Under this relation, a mass ratio of $q = 0.5$ corresponds to a brightness offset of $\Delta m \approx -0.107$ mag; binary candidates were therefore identified as stars displaced above the fitted single-star sequence by more than this threshold. Figure~\ref{fig:berk32_binary}a shows the CMD of the selected MS stars together with the fitted single-star sequence and the binary selection limits. Most stars are concentrated around the fitted MS locus, while a noticeable group of brighter stars forms a parallel sequence above it, consistent with the expected distribution of unresolved binaries. A broader spread is seen towards fainter magnitudes, likely due to increasing photometric uncertainties. The distribution of $\Delta G$ values is presented in Figure~\ref{fig:berk32_binary}b. The main peak is centered close to $\Delta G \approx 0$ mag, representing stars consistent with the single-star sequence. An additional excess of stars is visible below the adopted lower limit, suggesting the presence of a significant unresolved binary population. The cumulative distribution of $\Delta G$ is shown in Figure~\ref{fig:berk32_binary}c, where the binary fraction is indicated by the interval between the two selection limits. Among the 822 MS stars, 369 satisfy the binary criterion, giving a binary fraction of $f_{\rm b} = 0.449 \pm 0.017$, where the corresponding 1$\sigma$ confidence interval is $0.432 \leq f_{\rm b} \leq 0.466$. This value is comparable to binary fractions reported for other old OCs analysed with similar methods \citep[e.g.,][]{Milone2012, Donada2023, Cinar2026a}. 

The distribution of binary candidates across mass-ratio bins reveals 64 systems in the $0.5 \leq q < 0.6$ range, 90 in $0.6 < q \leq 0.7$, 82 in $0.7 < q \leq 0.8$, 93 in $0.8 < q \leq 0.9$, and 42 in $0.9 < q \leq 1.0$. The distribution is non-uniform, exhibiting a pronounced peak in the $0.8 < q \leq 0.9$ bin and a secondary enhancement around $0.6 < q \leq 0.7$. These features may reflect the preferential accumulation of detached systems near $\langle q\rangle \approx 0.88$, while the enhancement around $\langle q\rangle \approx 0.60$ may be associated with semi-detached binaries as well as possible contributions from contact and other interacting binary systems, rather than representing a uniquely defined binary class, as reported for field binary populations \citep[e.g.,][]{Bilir2005, Demircan2006, Eker2006, Ibanoglu2006, Eker2014}. Therefore, the observed excess near $\langle q\rangle \approx 0.60$ should be interpreted cautiously, since it likely reflects a mixed population of interacting binaries with overlapping mass-ratio characteristics. If confirmed by spectroscopic follow-up, this would suggest that the internal dynamical processes governing mass-ratio evolution in Berkeley~32 are broadly consistent with those seen in the general Galactic binary population, despite the cluster's old age and metal-poor composition. Since the adopted method is primarily sensitive to binaries with $q \geq 0.5$, the derived fraction should be considered a lower limit to the cluster's total binary population.

We also investigated the Blue Straggler Star (BSS) and variable-star populations of Berkeley~32 using cluster members with membership probabilities greater than 70\%. As expected for an old and dynamically evolved OC, Berkeley~32 exhibits a noticeable BSS population \citep{Linck2026}. To identify previously reported BSS candidates, we cross-matched our member sample with the catalogs of \citet{rain2021new} and \citet{Jadhav2021bss}. We recovered 22 of the 27 stars listed by \citet{rain2021new}, along with two Yellow Straggler Star (YSS) candidates. From the catalog of \citet{Jadhav2021bss}, we identified 17 out of 20 stars. In total, 17 stars are common to both the literature catalogs and our member sample, indicating good agreement with previous studies. In addition to the literature sources, we identified 48 photometric BSS candidates from their locations in the CMD, as shown in the right panel of Figure~\ref{fig:CMDs}. Although the membership probabilities are generally high, spectroscopic follow-up observations would be useful to assess potential contamination from field stars and unresolved binaries. The full list of BSS and YSS candidates is presented in Appendix Table~\ref{tab:appendix_bss}.

We also investigated the variable star content of Berkeley~32 by cross-matching our high-probability member sample ($P_{\rm mem} \geq 0.7$) with the Gaia DR3 variability catalog \citep[VizieR I/358/vclassre;][]{Gaia_Variability}. Using a matching radius of $1^{\prime\prime}$, we identified 11 variable stars associated with the cluster. Most of the identified variables are pulsating stars. Eight sources are classified as DSCT$|$GDOR$|$SXPHE variables, with $G$-band magnitudes ranging from 14.25 to 16.48 mag. Their effective temperatures range from approximately 7400 K to 9600 K, consistent with main-sequence pulsators. In addition, two RS-type variables were identified in the $G$-band magnitude range of 14.64--15.97 mag, corresponding to chromospherically active binary systems. One eclipsing binary (ECL) candidate was also detected at $G \simeq 15.65$ mag. The membership reliability of the sample is high, with nearly 70\% of the variables having membership probabilities above $P_{\rm mem} \geq 0.95$. The full list of variable star candidates is presented in Appendix Table~\ref{tab:appendix_bss}.

The identified variable stars provide additional information about the stellar population and evolutionary properties of Berkeley~32. In particular, the presence of several $\delta$~Sct and $\gamma$~Dor candidates suggests that the cluster hosts a significant population of pulsating main-sequence stars, which may be suitable targets for future asteroseismic studies. The RS-type systems and the eclipsing binary candidate may also help constrain binary evolution and stellar activity in an intermediate-age OC environment. Together, the BSS and variable star populations further highlight Berkeley~32 as an important laboratory for studying stellar evolution.

We also inspected publicly available \textit{TESS} photometry \citep{Ricker2015} for selected sources in the Berkeley~32 field. Light curves were analysed using Lomb-Scargle (LS) periodograms to search for possible periodic variability. However, due to the large \textit{TESS} pixel scale (see Figure~\ref{fig:tess_example} left panel) and strong crowding in the cluster region, a detailed variability analysis was not possible. A brief description of the data, analysis procedure, and a representative example are provided in Appendix~\ref{app:tess}.

\section{Dynamical Orbital Parameters}\label{sec:orbit}

\subsection{Radial Velocity Determination}\label{sec:rv}

\begin{figure}
    \centering
    \includegraphics[width=1\linewidth]{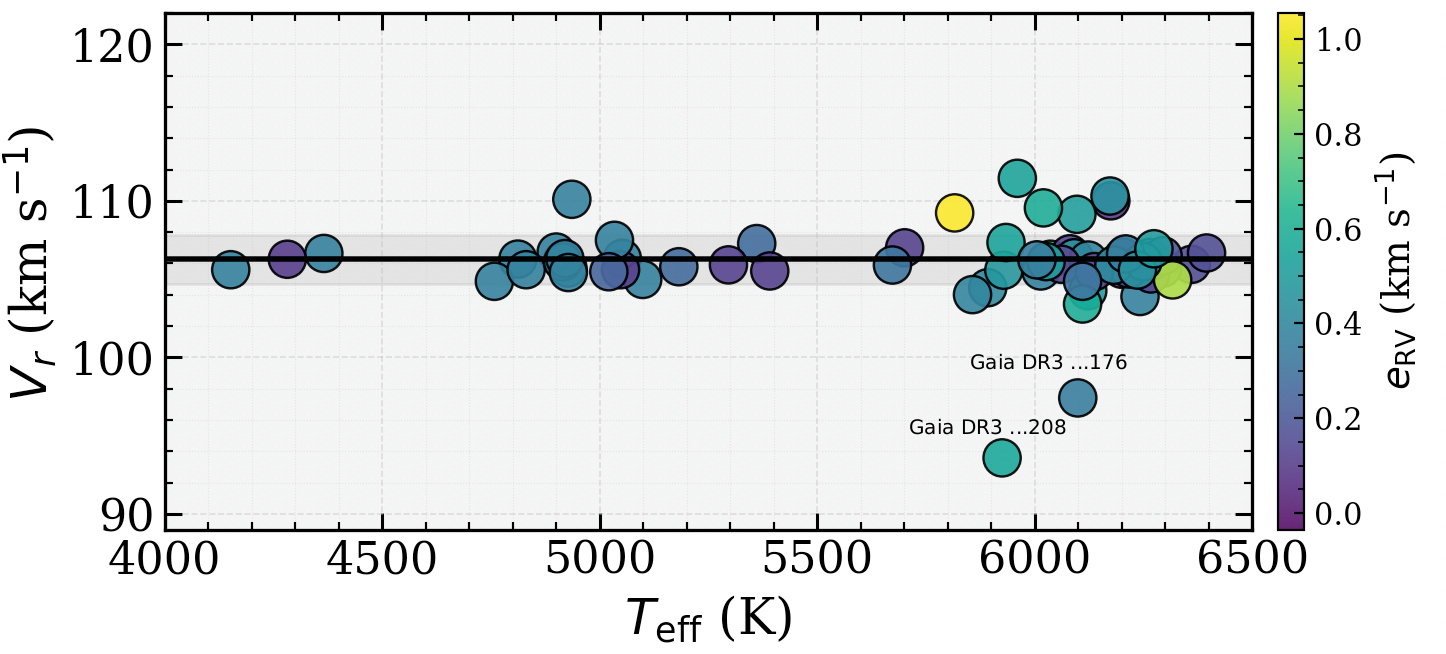}
    \caption{Distribution of radial velocities as a function of effective temperature for Berkeley~32 member stars. Points are color-coded according to the radial velocity error ($e_{\rm RV}$; km~s$^{-1}$). The solid horizontal line indicates the weighted mean cluster radial velocity, while the shaded region represents the corresponding $1\sigma$ internal velocity dispersion.}
    \label{fig:radial_vel}
\end{figure}

The mean radial velocity of Berkeley~32 was determined from spectroscopic data of the GES DR5.1 \citep{Hourihane2023}, which was preferred over available LAMOST and Gaia DR3 measurements due to its higher spectral resolution. Following \citet{Soubiran2018} and \citet{Carrera2022}, we computed the weighted mean radial velocity ($V_{\rm rad,OC}$) and internal velocity dispersion ($\sigma_{V_{\rm rad,OC}}$), assigning each star a weight $w_i = 1/\sigma_{v_{\rm rad},i}^2$. Based on a final sample of $N = 73$ member stars, we derived $V_{\rm rad,OC} = 106.26 \pm 0.03$~km~s$^{-1}$ (standard error of the mean) and $\sigma_{V_{\rm rad,OC}} = 1.26$~km~s$^{-1}$.

Figure~\ref{fig:radial_vel} shows the distribution of stellar radial velocities as a function of effective temperature for the final member sample. The majority of stars are tightly clustered around the mean cluster velocity across the full temperature range ($\sim$4000--6500~K), supporting the kinematic coherence of Berkeley~32. Two stars near $T_{\rm eff} \sim 6000$~K (Gaia DR3 sources 3129900143576658176 and 3129900487170766208) lie beyond the nominal $3\sigma$ interval, yet retain astrometric properties compatible with cluster membership, with $\mathrm{RUWE}$ values of 1.00 and 1.25 and CMD positions consistent with the MS locus. These stars likely represent candidate spectroscopic binaries whose line-of-sight velocities are temporarily shifted by orbital motion. Excluding them changes the cluster mean radial velocity by only $\sim$0.03~km~s$^{-1}$, well within the quoted uncertainties, and does not affect the kinematic conclusions of this study.

\subsection{Cluster Orbit Analysis}

We performed a comprehensive dynamical analysis of Berkeley~32 to trace its orbital history within the Milky Way. Reconstructing the orbital evolution of Berkeley 32 provides important constraints on its dynamical history, radial migration, and long-term interaction with the Galactic disk potential. The Galactic space motions and orbital integrations were carried out using the {\texttt galpy} package \citep{Bovy2015}. As a baseline model, we adopted the widely used axisymmetric {\sc MWPotential2014}, which represents the Galactic gravitational field through three main components: a Miyamoto--Nagai disk \citep{Miyamoto1975}, a Navarro--Frenk--White dark matter halo \citep{Navarro1996}, and a spherical bulge described by a power-law density profile with an exponential cut-off \citep{Bovy2013}. The total potential can be written as:
\begin{equation}
\Phi_{\rm total}(R, z) = \Phi_{\rm bulge}(r) + \Phi_{\rm disk}(R,z) + \Phi_{\rm halo}(r),
\end{equation}
where $r = \sqrt{R^2 + z^2}$ is the Galactocentric spherical radius. To assess the robustness of our orbital solutions against the choice of Galactic potential, we also integrated the orbits using an alternative model: the {\sc McMillan2017} potential \citep{McMillan2017}, which incorporates an updated mass model with revised disk and halo parameters based on a broad set of observational constraints. In all cases, the derived orbital parameters were consistent with those obtained from {\sc MWPotential2014} within the uncertainties. We therefore adopt this potential as our fiducial model, in line with numerous OCs studies in the literature \citep[e.g.,][]{Cantat-Gaudin20, Tarricq2021, Donor2020, Yontan2023c, Tasdemir23, Yucel24, Tasdemir2025, Cinar2025, Cinar2026}. 

While the axisymmetric model provides a robust first-order description of the Galactic potential, it neglects important non-axisymmetric structures known to influence stellar orbits. To account for these effects, we also incorporated rotating-bar and spiral-arm perturbations. The Galactic bar was modeled using the {\tt DehnenBarPotential}, which represents a rotating quadrupole perturbation of the form
\begin{equation}
\Phi_{\rm bar}(R, \phi, t) = A(t)\cos\left[2(\phi - \Omega_{\rm bar} t)\right] f(R),
\end{equation}
where $\Omega_{\rm bar}$ is the bar pattern speed and $A(t)$ describes its time-dependent amplitude. In addition, spiral structure was included via a steady-state spiral perturbation (e.g., {\tt SpiralArmsPotential}), allowing us to explore the combined impact of bar- and spiral-induced resonances on cluster orbits. For the non-axisymmetric components, we adopted typical literature values for the bar pattern speed ($\Omega_{\rm bar} \sim 40$ km\,s$^{-1}$\,kpc$^{-1}$), bar scale length ($R_{\rm bar} \sim 3.5$ kpc), and a modest perturbation amplitude, ensuring consistency with previous dynamical studies of the Milky Way.

The orbital integrations were performed using the full six-dimensional phase-space information: equatorial coordinates ($\alpha$, $\delta$), heliocentric distances ($d$), proper-motion components ($\mu_\alpha \cos\delta$, $\mu_\delta$), and systemic radial velocities (Section~\ref{sec:rv}). We adopt a local circular velocity of $V_{\rm rot} = 220$~km~s$^{-1}$ and a solar offset of $Z_0 = 25 \pm 5$~pc \citep{Bovy2012}, and compute the Galactocentric distance via the standard geometric transformation \citep{Tuncel2019}, yielding $R_0 = 8.2 \pm 0.1$~kpc. The guiding radius, defined as the radius of a circular orbit with the same specific angular momentum as the cluster, provides a useful diagnostic for radial migration and orbital mixing \citep[e.g.][]{Binney2008, Schoenrich2009}. Orbital trajectories were integrated both backward and forward in time over the derived cluster age with a timestep of 1~Myr.

\begin{figure}
    \centering
    \includegraphics[width=0.85\linewidth]{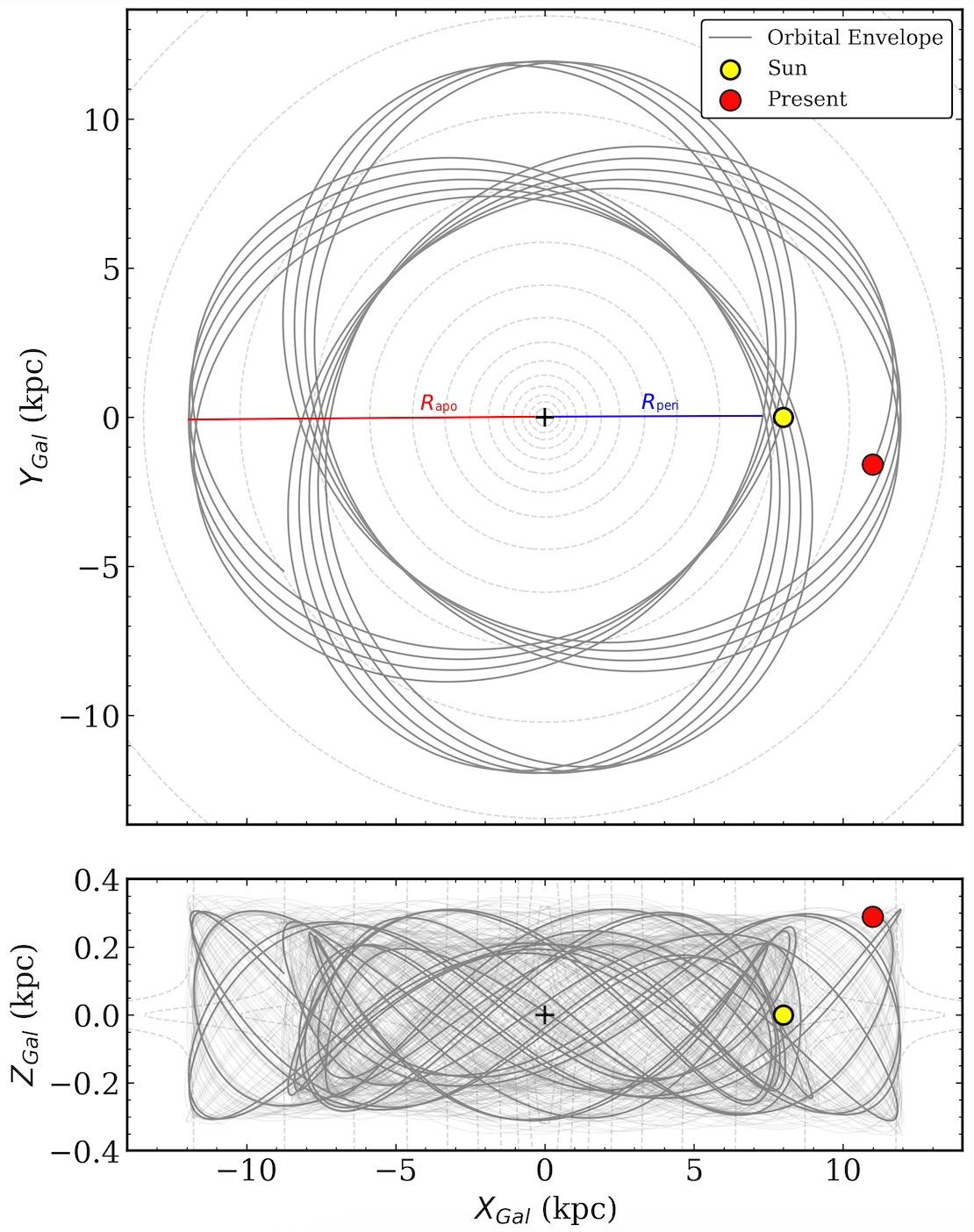}
    \caption{Galactic orbit of the cluster. \textit{Top:} Orbital envelope in the $X_{\mathrm{Gal}}$--$Y_{\mathrm{Gal}}$ plane. \textit{Bottom:} Orbit in the $X_{\mathrm{Gal}}$--$Z_{\mathrm{Gal}}$ plane, where light gray lines denote sampled orbits reflecting observational uncertainties. In both panels, the present-day position of the cluster is marked by a red-filled circle, the Sun's location is indicated by a yellow-filled circle, and the background gray contours illustrate the Galactic density.}
    \label{fig:orbit}
\end{figure}

To account for observational uncertainties, we performed a Monte Carlo analysis by generating multiple realizations of the input phase-space parameters. The resulting ensemble of orbital solutions yielded robust estimates of the apogalactic distance ($R_{\rm apo}$), perigalactic distance ($R_{\rm peri}$), mean orbital radius ($R_{\rm m}$), eccentricity ($e$), maximum vertical height ($Z_{\rm max}$), guiding radius ($R_{\rm g}$), and orbital period ($P_{\rm orb}$), all listed in Table~\ref{tab:summary}. Berkeley~32 follows a moderately eccentric orbit ($e = 0.268$) with $R_{\rm peri} = 6.911$~kpc and $R_{\rm apo} = 11.971$~kpc (Figure~\ref{fig:orbit}), and a mean orbital radius of $R_{\rm m} = 9.447$~kpc, indicating that the cluster spends a significant fraction of its time in the outer disk. The maximum vertical height of $Z_{\rm max} = 0.311$~kpc is consistent with thin-disk kinematics. The moderately eccentric orbit and large apogalactic distance support a scenario of substantial secular evolution likely driven by resonant interactions with the Galactic bar and spiral arms \citep[e.g.,][]{Onal2018}. We caution that long-term backward integrations remain sensitive to uncertainties in the Galactic potential and non-axisymmetric perturbations; the inferred birth radii should therefore be regarded as approximate indicators of past orbital evolution rather than precise formation locations.

\section{Dynamical State and Mass Segregation}\label{sec:relax}

OCs are relatively low-density and weakly bound stellar systems that evolve under the combined influence of internal stellar encounters and external Galactic tidal effects \citep{Inagaki85, Baumgardt03}. Over time, repeated two-body interactions redistribute kinetic energy among cluster members, gradually driving the system toward a state of partial energy equipartition and dynamical equilibrium \citep{Spitzer71, Binney2008}. One of the most prominent consequences of this long-term dynamical evolution is mass segregation, in which massive stars preferentially migrate toward the cluster center while lower-mass stars populate the outer regions \citep{Dib18, Bisht20}. Such behaviour has been identified in numerous Galactic OCs spanning a wide range of ages and masses \citep{Joshi20, Tanik2025, Karagoz25}, indicating that mass segregation is a common outcome of cluster dynamical evolution. The characteristic timescale associated with this process is the dynamical relaxation time, $T_{\rm relax}$, which describes the time required for stellar encounters to significantly alter the velocity distribution of cluster members. Following \citet{Spitzer71}:
\begin{equation}
\label{eq:Trelax}
T_{\rm relax} = \frac{8.9\times10^5\,\sqrt{N}\,R_h^{3/2}}{\sqrt{\overline{M_C}}\,\log(0.4\,N)},
\end{equation}
where $N$ is the number of members, $\overline{M_C}$ is the mean stellar mass, and $R_h$ is the half-mass radius. We estimated $R_h$ using \citet{Sableviciute_2006}:
\begin{equation}
R_{\rm h} = 0.547\,r_{\rm c} \left(\frac{r_{\rm t}}{r_{\rm c}}\right)^{0.486},
\end{equation}
where $r_{\rm c}$ and $r_{\rm t}$ are the core and tidal radii. The derived values are $r_{\rm c} = 1.39 \pm 0.08$~pc and $r_{\rm t} = 9.65 \pm 3.11$~pc for Berkeley~32, giving $R_{\rm h} = 1.95$~pc. Using the derived structural parameters, we obtain a dynamical relaxation time of approximately 30 Myr, which is significantly shorter than the cluster age.

The dynamical age parameter $\tau = \mathrm{age}/T_{\rm relax}$ \citep{Spitzer71, Binney2008}, previously applied to OCs by \citet{2021JApA...42...90E, 2022JApA...43...26E}, describes the dynamical state. We find $\tau = 164$ for Berkeley~32, indicating that Berkeley~32 is dynamically relaxed. Taken together with the orbital results, this suggests that Berkeley~32 has undergone both internal evolution and radial migration. Detailed analyses of the luminosity and mass function in Appendix~\ref{sec:mf}. Mass segregation is one of the key diagnostics of the dynamical evolution of stellar clusters. It arises from two-body relaxation, through which kinetic energy is redistributed among cluster members, driving the system toward partial energy equipartition. As a result, more massive stars progressively lose kinetic energy and migrate toward the cluster center, while lower-mass stars gain kinetic energy and are displaced to larger radii \citep{PortegiesZwart2010, Dib2018}. Given the relatively large distance of Berkeley~32, the photometry becomes incomplete at the faint end of the main sequence, limiting the reliable sampling of low-mass stars. Therefore, instead of using the full luminosity (or mass) range, we perform the mass segregation analysis on well-defined, completeness-limited stellar populations across different magnitude intervals, ensuring a homogeneous comparison between radial bins.

\begin{figure}[ht]
    \centering
    \includegraphics[width=0.9\linewidth]{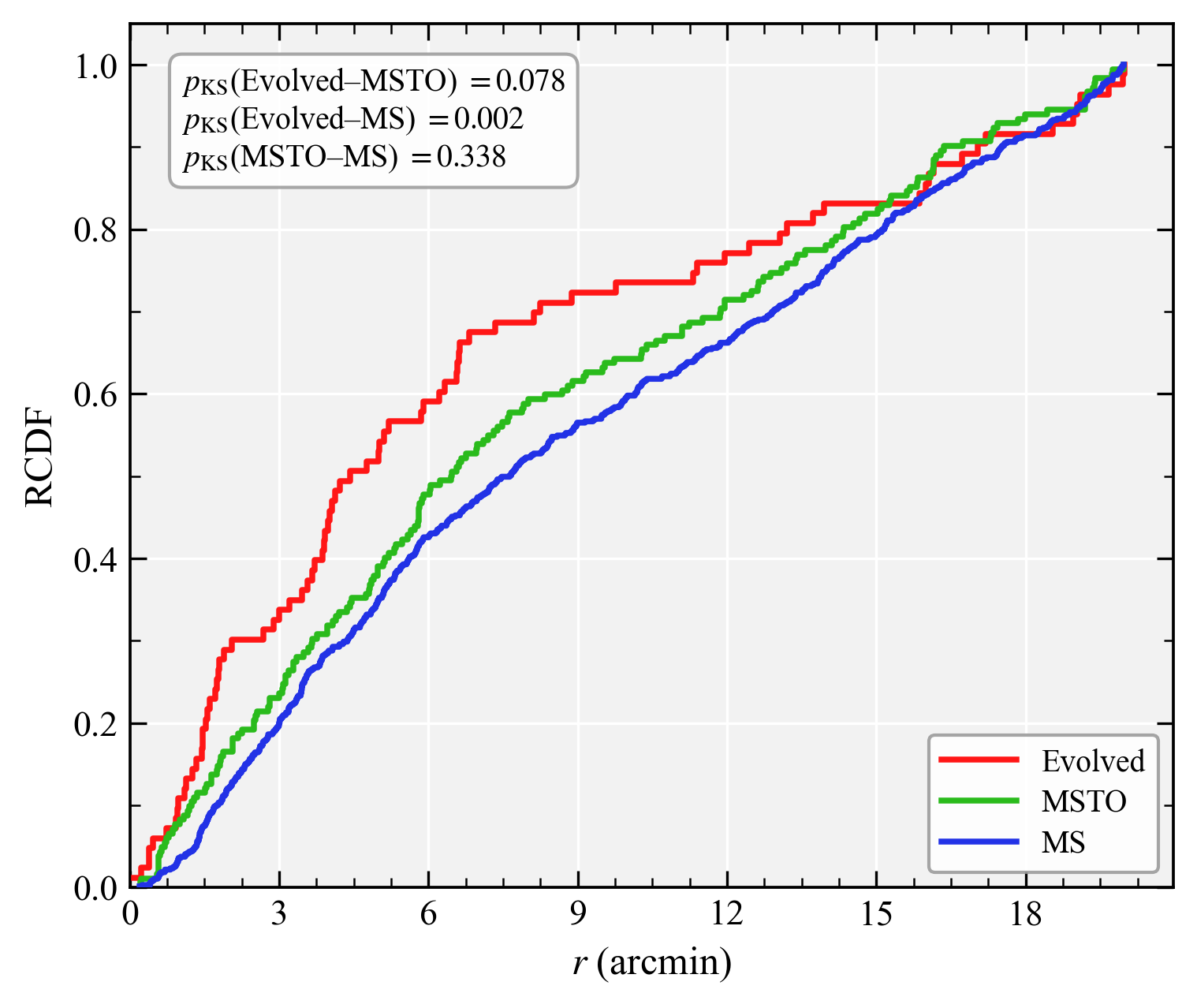}
    \caption{RCDFs of evolved, MSTO, and MS populations in Berkeley~32. The radial distance corresponds to the projected angular separation from the cluster center in arcminutes. Populations were selected according to their $G$-band magnitude ranges and membership probabilities.}
    \label{fig:rcdf_berk32}
\end{figure}

Cluster members were divided into three classes based on their CMD positions: evolved (red-giant) stars, MSTO stars, and MS stars, which serve as proxies for decreasing stellar mass. The spatial distributions were examined using the RCDFs shown in Figure~\ref{fig:rcdf_berk32}. The evolved ($N = 83$) and MSTO ($N = 182$) stars exhibit stronger central concentration than the MS stars ($N = 639$): the evolved population dominates within $\sim$6~arcmin of the cluster center, MSTO stars are more prominent between $\sim$6 and $\sim$15~arcmin, and the MS population increasingly dominates beyond $\sim$9~arcmin, indicating that lower-mass stars are more widely distributed in the outer regions. To quantify these spatial differences, we applied two-sample Kolmogorov--Smirnov (KS) tests to each pair of radial distributions. The evolved--MS comparison yields $p = 0.002$, confirming statistically distinct distributions, while the evolved--MSTO comparison gives $p = 0.078$, a marginal difference likely reflecting the small evolved sample size ($N = 83$). The MSTO--MS comparison shows no significant difference ($p = 0.338$). Together, the KS results confirm that the most massive (evolved) stars are significantly more centrally concentrated than the MS population, with MSTO stars occupying an intermediate distribution, providing clear evidence of mass segregation driven by long-term internal relaxation.

\section{Discussion}\label{sec:discussion}

\subsection{Comparison of Astrophysical Parameters with Previous Studies}

\begin{figure*}
    \centering
    \includegraphics[width=0.90\linewidth]{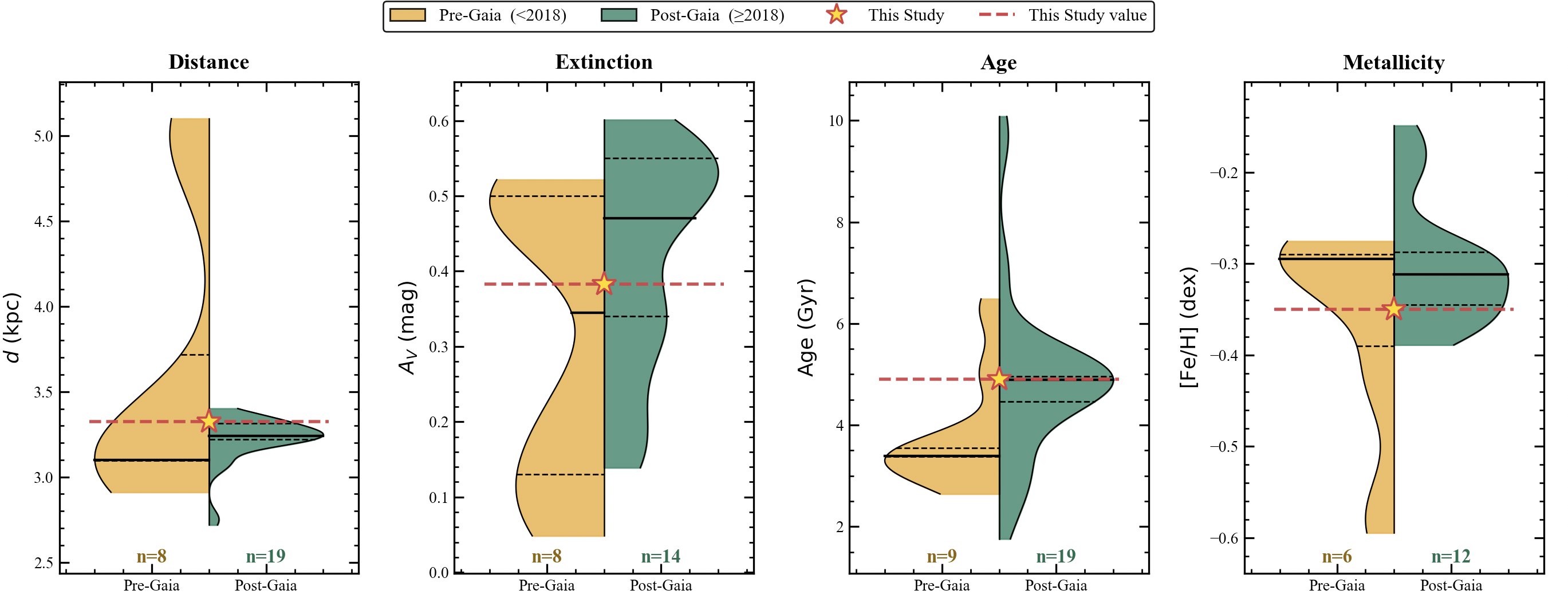}
    \caption{Literature values for Berkeley~32 before and after the \textit{Gaia} era. The panels show the distributions of distance, extinction ($A_V$), age, and metallicity ([Fe/H]), with pre-\textit{Gaia} studies ($\tau< 2018$; orange) and post-\textit{Gaia} studies ($\tau\geq 2018$; green). The dashed black lines mark the medians. The red star shows our result, and the horizontal red dashed line highlights its value. Sample sizes are given at the bottom of each panel.}
    \label{fig:lit_comp}
\end{figure*}

Figure~\ref{fig:lit_comp} and Table~\ref{tab:lit_comp} summarise the published determinations of the fundamental parameters of Berkeley~32 before and after the \textit{Gaia} era. This compilation provides a historical overview of published determinations rather than a weighted comparison of measurement precision. Pre-\textit{Gaia} studies show a considerably larger scatter, particularly in distance and age, reflecting the limitations of earlier astrometric and photometric data. With the advent of \textit{Gaia}, the published values have converged to a much narrower range across all four parameters, providing a more consistent picture of the cluster properties. Our derived values ($\langle d\rangle=3325$~pc, $\langle A_V\rangle=0.38$~mag, $\langle\tau\rangle=4.9$~Gyr, and $\langle\mathrm{[Fe/H]}\rangle=-0.39$~dex), marked by the red star in each panel, are fully consistent with this modern \textit{Gaia}-based consensus. The improvement is particularly evident in the distance and age estimates, while the extinction and metallicity distributions are also more tightly constrained in recent studies.

For the radial velocity, our weighted mean $V_{\rm rad,OC}=106.26\pm0.03$ km~s$^{-1}$ (73 members, GES DR5.1) is in excellent agreement with \citet{DOrazi2006} ($106.7\pm8.5$ km~s$^{-1}$) and \citet{Randich2009}  ($105.20\pm0.86$ km~s$^{-1}$), while the substantially reduced uncertainty reflects the precision of the GES data and the robustness of our membership selection.

The literature compilation intentionally includes all published determinations, irrespective of the observational technique employed, to provide a comprehensive historical overview of the parameter estimates for Berkeley~32. Consequently, no quality-based weighting or selection among different methods was applied in this comparison.

\begin{table}
\centering
\caption{Comparison of astrophysical parameters of Berkeley~32 from the literature. Dashes indicate parameters not reported.}
\label{tab:lit_comp}
\begin{tabular}{cccccc}
\hline\hline
Year & $d$ & $A_{\rm V}$ & Age & $[\mathrm{Fe/H}]$ & Ref. \\
 & (kpc)& (mag) & (Myr)& (dex) & \\
\hline
2026 & --    & --   & 4898 & $-0.34$ & [01] \\
2025 & --    & 0.34 & 4900 & $-0.29$ & [02] \\
2025 & 3.35  & 0.58 & 4764 & $-0.36$ & [03] \\
2024 & 3.22  & 0.16 & 3252 & --      & [04] \\
2024 & 3.20  & 0.25 & 7026 & --      & [05] \\
2022 & --    & --   & --   & $-0.38$ & [06] \\
2022 & 3.31  & --   & 5248 & $-0.30$ & [07] \\
2021 & 3.32  & 0.55 & 4853 & $-0.28$ & [08] \\
2021 & 3.37  & --   & 4898 & $-0.38$ & [09] \\
2020 & 3.08  & 0.34 & 4898 & --      & [10] \\
2020 & 3.23  & 0.55 & 2138 & --      & [11] \\
2020 & 3.24  & 0.47 & 9700 & $-0.34$ & [12] \\
2019 & --    & --   & 4170 & $-0.25$ & [13] \\
2017 & 3.01  & 0.15 & 6310 & --           & [14] \\
2016 & 5.00  & 0.07 & 2818 & --           & [15] \\
2016 & 11.19 & --   & 3850 & $-0.30$  & [16] \\
2013 & 5.00  & 0.07 & 2818 & $-0.29$     & [17] \\
2012 & 3.10  & --   & 3388 & $-0.29$     & [18] \\
2002 & 3.10  & 0.50 & 3388 & $-0.58$     & [19] \\
\hline
\end{tabular}
\begin{minipage}{\linewidth}
\footnotesize
\tablenotetext{}{
\textbf{References:}
[01]~\citet{Otto2026},
[02]~\citet{Bijavara2025},
[03]~\citet{Li2025},
[04]~\citet{Hunt2024},
[05]~\citet{Rain2024},
[06]~\citet{Spina2022},
[07]~\citet{Netopil2022},
[08]~\citet{Dias2021},
[09]~\citet{Spina2021},
[10]~\citet{Cantat-Gaudin2020},
[11]~\citet{Kounkel2020},
[12]~\citet{Zhong2020},
[13]~\citet{Liu2019},
[14]~\citet{Loktin2017},
[15]~\citet{Kharchenko2016};
[16]~\citet{Netopil2016},
[17]~\citet{Kharchenko2013},
[18]~\citet{Gozha2012},
[19]~\citet{Dias2002}.}
\end{minipage}
\end{table}

\subsection{The [Y/Mg] Chemical Clock}\label{sec:Ymg_age}

As an independent check on the isochrone age, we exploited the [Y/Mg] chemical clock. Abundance ratios between neutron-capture and $\alpha$-elements provide a practical empirical age indicator for Galactic stellar populations \citep{Masseron2015}: Mg is synthesized rapidly in core-collapse supernovae, whereas Y is produced via the slow neutron-capture (s-)process in AGB stars on longer timescales \citep[e.g.,][]{Nissen2015, DelgadoMena2017}. As a result, [Y/Mg] systematically decreases with increasing stellar age, providing a chronometric diagnostic independent of photometric fitting.

Nevertheless, the [Y/Mg]--age relation is not strictly universal across the Galactic disk. The efficiency of the s-process in AGB stars is highly sensitive to initial metallicity, and different regions of the Galaxy have experienced distinct star formation histories \citep{Spina2018}. The relation also becomes less tight at lower metallicities and for evolved stars \citep{Feltzing2017}. To account for this metallicity dependence, we adopted the multivariate calibration of \citet{Casali2020}, which incorporates [Fe/H] as a secondary variable:
\begin{equation}
\text{Age (Gyr)} = 5.245 + 5.057 \times \text{[Fe/H]} - 32.546 \times \text{[Y/Mg]}.
\label{eq:Y_Mg}
\end{equation}
To apply this calibration to our sample, we performed a strict star-by-star analysis rather than using mean cluster abundances. First, the absolute abundances of Y and Mg derived for each star were normalized to the solar values from \citet{Asplund2009} ($A(\text{Y})_{\odot} = 2.21$ dex and $A(\text{Mg})_{\odot} = 7.60$ dex) to calculate [Y/H] and [Mg/H]. The [Y/Mg] ratio was then computed directly from these normalized abundances ($\text{[Y/Mg]} = \text{[Y/H]} - \text{[Mg/H]}$). Finally, the chemical age of each star was determined by evaluating Equation \ref{eq:Y_Mg} using the star's specific [Fe/H] and [Y/Mg] values.

By calculating the statistical mean of these individual stellar ages, we obtain a mean chemical age of $4.73 \pm 2.39$~Gyr for the 17-member stars of Berkeley~32. This is in good agreement with the isochrone age of $4.9\pm 0.5$~Gyr derived above, as well as with ages reported in the literature for this cluster. The relatively large uncertainty reflects the intrinsic star-to-star abundance scatter within the cluster, a known limitation of chemical clocks applied to giant stars. Also, when adopting the cluster mean abundance values of $\langle {\rm [Y/Mg]} \rangle = -0.03$ dex and $\langle {\rm [Fe/H]}\rangle = -0.39$ dex, the corresponding chemical age is estimated to be 4.25~Gyr. We thus treat the [Y/Mg] age as a complementary diagnostic rather than a primary constraint, consistent with the recommendation of \citet{Casali2020} that such calibrations are most reliable for stellar populations older than $\sim$1~Gyr.

\subsection{Chemical Birth Radii and Radial Migration}

\begin{figure*}
    \centering
    \includegraphics[width=0.55\linewidth]{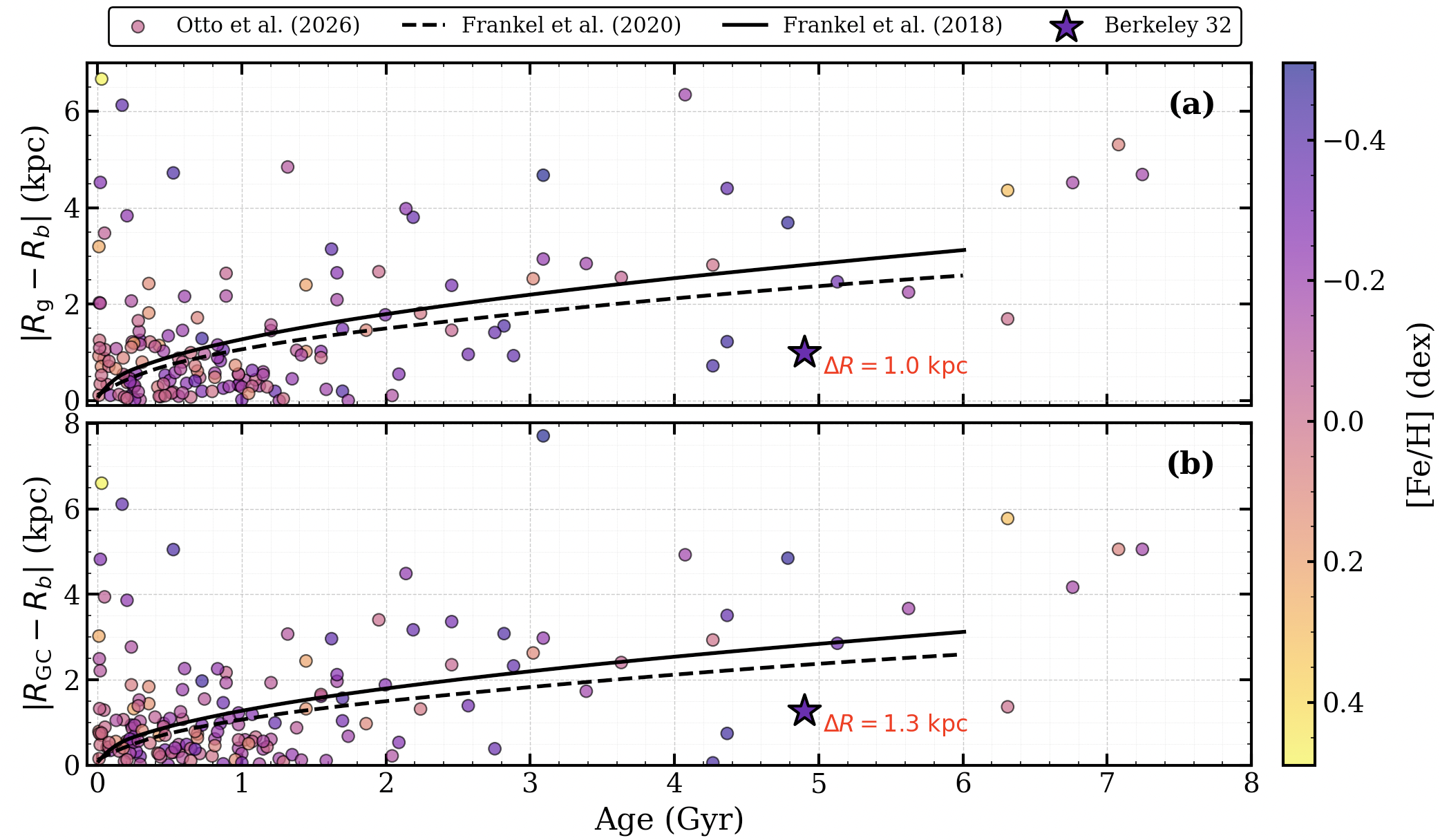}
   \vspace{-0.2cm}
    \caption{Distribution of OCs in the plane of age versus the absolute difference between the guiding radius and the present Galactocentric radius ($|R_{\rm g} - R_b|$; panel a) and between the birth radius and the present Galactocentric radius ($|R_{\rm GC} - R_b|$; panel b). The color scale represents the cluster metallicity ([Fe/H]) in dex. The background circular data points are taken from \citet{Otto2026}, while Berkeley~32 is highlighted with a star symbol. The black solid and dashed lines indicate empirical upper limits for age-dependent kinematic heating and radial epicyclic amplitude from \citet{Frankel_2018} and \citet{Frankel2020}, respectively.}
    \label{fig:r_guiding}
\end{figure*}

The chemical birth radius ($R_{\rm b}$) of Berkeley~32 was estimated following the chemo-dynamical approach of \citet{Minchev2018}, as adopted in subsequent studies \citep[e.g.,][]{Lu2024, Ratcliffe2025, Ratcliffe2026}. In this framework, $R_{\rm b}$ is inferred from the cluster's age and metallicity under the assumption of a time-dependent Galactic radial metallicity gradient. Using $\rm [Fe/H] = -0.39$~dex and an age of $\approx 4.9$~Gyr, we obtain $R_{\rm b} = 9.82$~kpc, somewhat larger than the solar radius ($R_\odot \approx 8.2$~kpc), suggesting an outer-disk origin. This estimate depends on the adopted chemo-dynamical model and should be regarded as approximate rather than a unique solution. Differences between $R_{\rm b}$ and $R_{\rm g}$ are interpreted as signatures of churning, long-term changes in the guiding-center radius driven by angular momentum exchange with spiral arms and the Galactic bar \citep{Sellwood2002, Anders2017}, whereas offsets between $R_{\rm g}$ and $R_{\rm GC}$ reflect epicyclic motion (blurring) arising from orbital eccentricity \citep{Binney2008, Sellwood2002}. For Berkeley~32, $|R_{\rm g} - R_{\rm GC}| \approx 2.3$~kpc indicates a measurable contribution from blurring, while $|R_{\rm g} - R_{\rm b}| \approx 1$~kpc is consistent with moderate radial migration through churning.

In Figure~\ref{fig:r_guiding}, Berkeley~32 is shown in the age--radius deviation plane together with the OC sample analysed by \citet{Otto2026}. With an age of $\sim4.9$~Gyr, the cluster exhibits $|R_{\rm g} - R_{\rm b}| \approx 1$~kpc (Figure~\ref{fig:r_guiding}a), tracing the churning component, while $|R_{\rm GC} - R_{\rm b}| \approx 1.3$~kpc (Figure~\ref{fig:r_guiding}b) reflects the combined effects of churning and blurring. The latter quantity is qualitatively consistent with the additional contribution expected from epicyclic motion, as inferred from the orbital analysis. Both quantities fall below the empirical secular heating envelopes defined by \citet{Frankel_2018} and \citet{Frankel2020}, indicating that the present-day kinematics of Berkeley~32 are compatible with secular dynamical heating expected for a cluster of its age. No anomalous perturbation is therefore required to explain its current orbital properties, although moderate radial migration induced by interactions with spiral structure or giant molecular clouds remains plausible. In contrast, several young OCs ($t < 0.5$~Gyr) in the same sample (see Figure~\ref{fig:r_guiding}a) exhibit extremely large $|R_{\rm guiding}-R_{\rm birth}|$ values reaching up to $\sim 6.5$~kpc. Given the short dynamical timescales involved, such large displacements are difficult to reconcile with standard radial migration scenarios. Therefore, these extreme values may instead reflect the number of member stars, uncertainties in the estimated birth radii, metallicity gradients, or orbital parameters.

\begin{figure}
    \centering
    \includegraphics[width=1\linewidth]{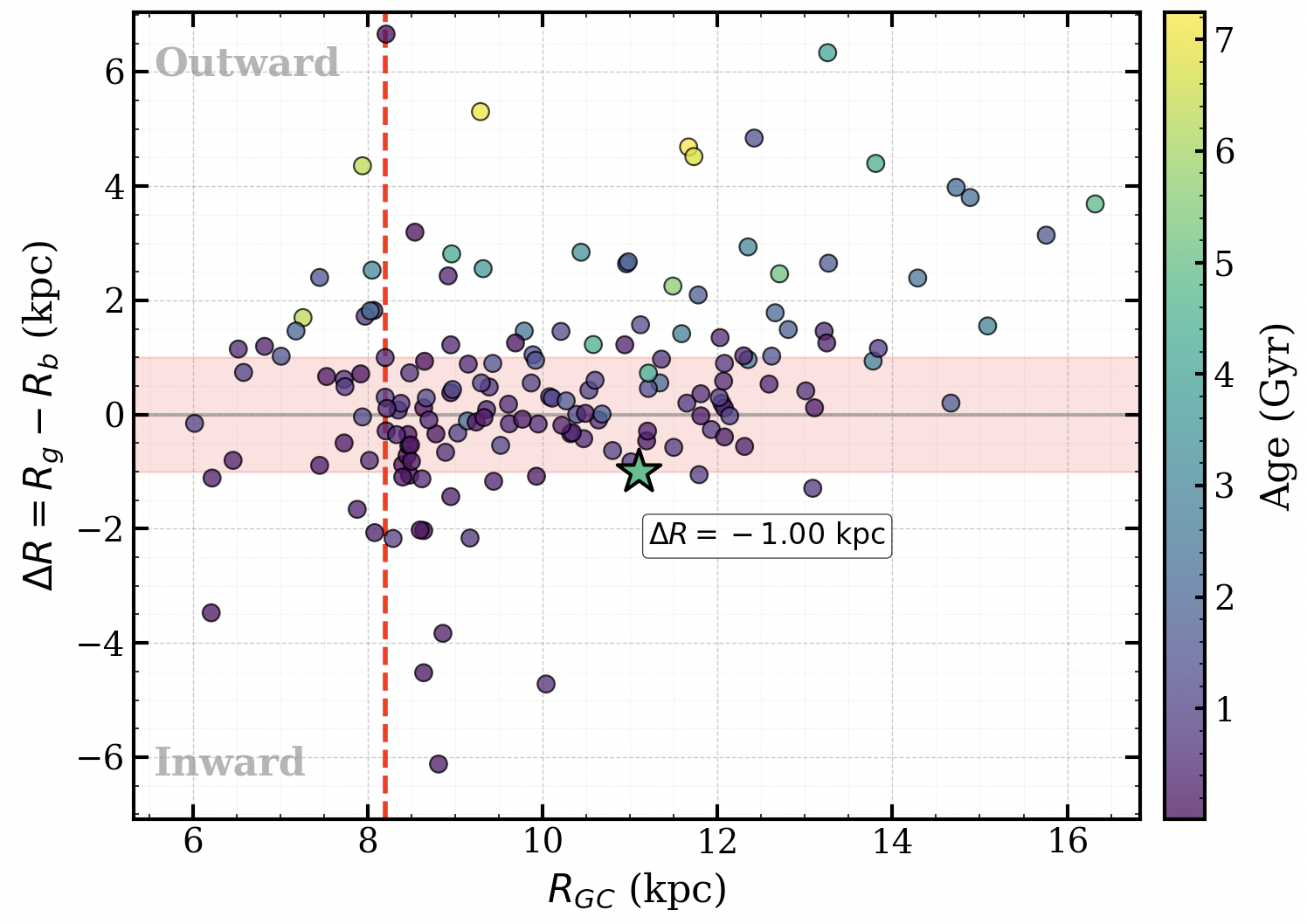}
    \caption{Radial migration distance $\Delta R = R_{\rm g} - R_{\rm b}$ as a function of present Galactocentric radius $R_{\rm GC}$ for OCs from \citet{Otto2026} (circles) and Berkeley~32 (star symbol). Points above the grey horizontal line ($\Delta R > 0$) indicate outward migrators, while those below ($\Delta R < 0$) indicate inward migrators. The red dashed vertical line marks the solar Galactocentric radius ($R_{\rm GC},_\odot \approx 8.2$~kpc). Clusters are color-coded by age. The shaded red horizontal band highlights the region of minimal radial migration ($-1 \le \Delta R \le 1$~kpc).}
    \label{fig:Delta_R}
\end{figure}

Within the \citet{Donada2026} sample, most inward migrators exhibit $|R_{\rm g} - R_{\rm b}|$ below $\sim$1~kpc, whereas Berkeley~32 yields $|R_{\rm g} - R_{\rm b}| \approx 1$~kpc, placing it among moderate churning candidates following the threshold of \citet{Netopil2022}. This is notable because old OCs in the \citet{Otto2026} sample predominantly migrate outward (38 out of 39 clusters with $|\Delta R| > 1$~kpc), while Berkeley~32 shows a moderate inward displacement ($\Delta R \approx -1$~kpc), consistent with secular angular momentum redistribution \citep{Sellwood2002, Chen2020}. Overall, 54\% of clusters in the \citet{Otto2026} sample exhibit $|\Delta R| < 1$~kpc, suggesting that strong churning is not significant for the majority of the population, and that dominant migration mechanisms differ between young and old OC populations owing to differences in their Galactic environments and long-term dynamical evolution.

For the older OC population, applying the prescription of \citet{Chen2020} yields an estimated migration rate of $1.0 \pm 0.6~\rm{kpc~Gyr}^{-1}$ based on the 39 old OCs with $|\Delta R| > 1$~kpc in the \citet{Otto2026} sample. Although somewhat lower than the value of $1.5 \pm 0.5~\rm{kpc~Gyr}^{-1}$ reported by \citet{Chen2020}, this estimate remains compatible with secular radial mixing operating over several Gyrs within the Galactic disk. In this context, Berkeley~32 appears particularly noteworthy because, despite its relatively old age ($\sim4.9$~Gyr), it exhibits $\Delta R \approx -1$~kpc at $R_{\rm GC} \approx 11.11$~kpc, placing it among the inward migrators in Figure~\ref{fig:Delta_R}. This behaviour contrasts with the dominant outward migration trend observed among older OCs in the \citet{Otto2026} sample and may indicate a more complex chemo-dynamical evolutionary history involving secular redistribution of angular momentum.

\begin{figure}
    \centering
    \includegraphics[width=1\linewidth]{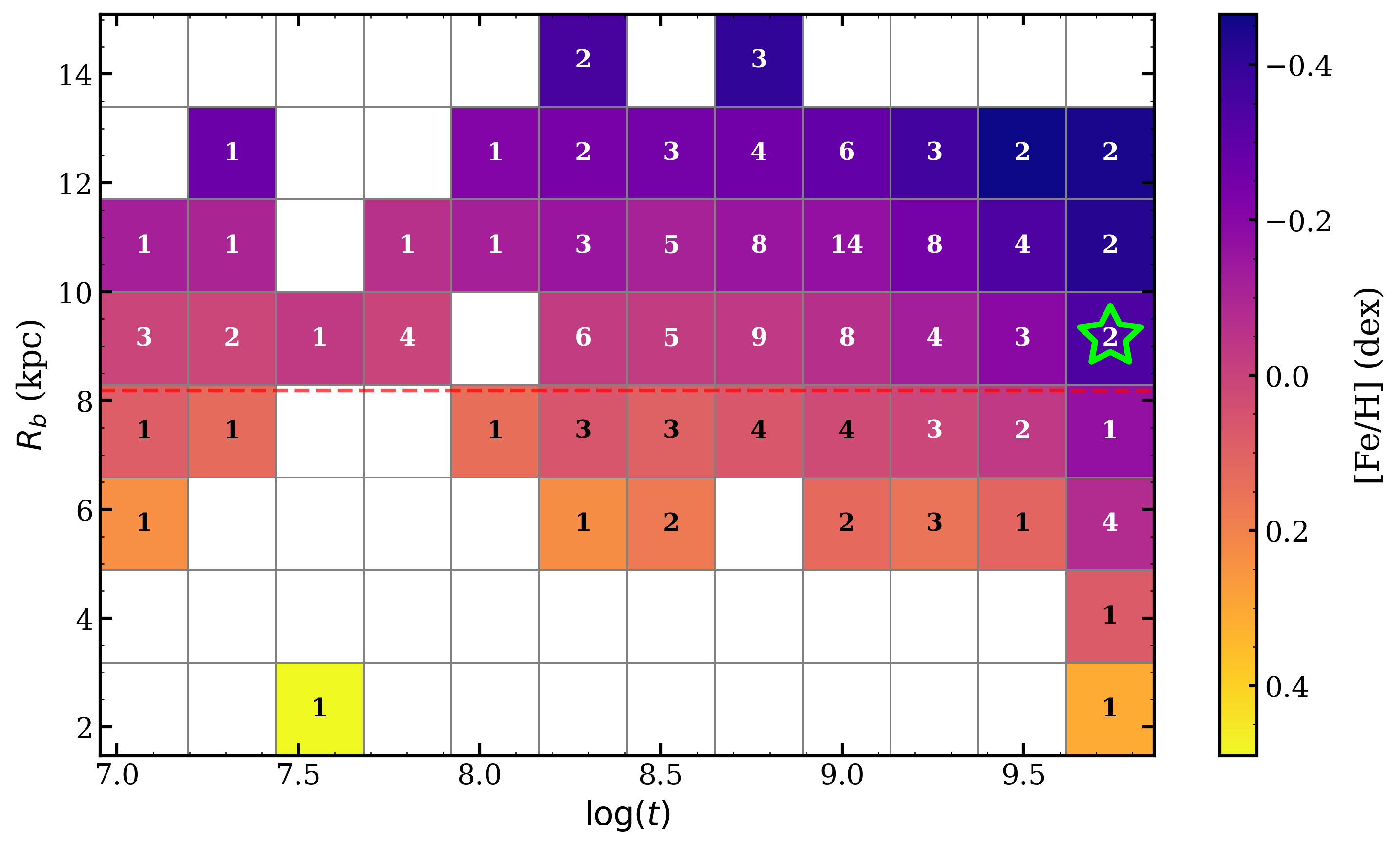}
    \caption{Distribution of OCs from \citet{Otto2026} in the $\log(t)$--$R_{\rm b}$ plane. Each cell is color-coded by mean [Fe/H] (dex), with the number of clusters indicated. The light green star symbol represents Berkeley 32. The red dashed line indicates the Sun’s distance from the Galactic center ($R_{\rm GC},_\odot \approx 8.2$~kpc).}
    \label{fig:rb_metallicity}
\end{figure}

Figure~\ref{fig:rb_metallicity} shows the distribution of clusters from \citet{Otto2026} in the $\log(t)$--$R_{\rm b}$ plane, color-coded by mean metallicity. A clear negative radial metallicity gradient is apparent — clusters at smaller $R_{\rm b}$ tend to be more metal-rich, while those at larger radii are more metal-poor, consistent with inside-out disk formation \citep[e.g.,][]{Chiappini1997, Minchev2014}. Berkeley~32, with $\log(t) = 9.69$, $R_{\rm b} = 9.82$~kpc, and $\langle\rm [Fe/H]\rangle = -0.39 \pm 0.02$~dex, lies in a region populated by old, metal-poor clusters beyond the solar circle, consistent with its inferred outer-disk origin and moderate inward radial migration. Berkeley~32, with $\log(t) = 9.69$, $R_{\rm b} = 9.82$~kpc, and $\Delta R \approx -1$~kpc, is a moderate inward migrator, in contrast to the dominant outward migration trend among old OCs in the \citet{Otto2026} sample, yet consistent with secular angular momentum redistribution driven by spiral arms or the Galactic bar \citep{Sellwood2002}. The agreement between its birth radius, present-day orbital properties, and chemical abundances supports a scenario of outer-disk formation followed by moderate radial migration and epicyclic motion, while retaining the chemical characteristics of its birth environment. Its survival over $\sim$5~Gyr and well-constrained chemo-dynamical properties establish Berkeley~32 as a valuable benchmark for probing the interplay between radial migration and chemical evolution in the Galactic disk.

\section{Summary and Conclusion}
\label{sec:conclusion}

We present a multi-dimensional analysis of the old OC Berkeley~32, combining Gaia DR3 astrometry with GES spectroscopic abundances. The aim is to constrain its fundamental parameters, dynamical properties, and chemical birth radius, and to place these results in the context of Galactic disk evolution. The main findings can be summarized as follows:

\begin{enumerate}
\item Cluster membership was determined using a GMM applied to \textit{Gaia} DR3 proper motions and parallaxes. The resulting member sample is broadly consistent with previous catalogs \citep{Cantat-Gaudin2020, Hunt2024}. The mean proper motions ($\langle\mu_{\alpha}\cos\delta\rangle = -0.355 \pm 0.169$~mas~yr$^{-1}$, $\langle\mu_{\delta}\rangle = -1.589 \pm 0.154$~mas~yr$^{-1}$) and mean trigonometric parallax ($\langle\varpi\rangle = 0.273 \pm 0.072$~mas) agree with earlier estimates within uncertainties.

\item A King profile fit to the radial density distribution gives a core radius of $r_{\rm c} = 1.30 \pm 0.13$~arcmin and a tidal radius of $r_{\rm t} = 9.06 \pm 0.24$~arcmin, corresponding to $r_{\rm c} = 1.39 \pm 0.08$~pc and $r_{\rm t} = 9.65 \pm 3.11$~pc. The concentration parameter ($C \approx 0.84$) suggests a moderately concentrated and dynamically evolved structure.

\item Isochrone fitting to the Gaia CMD using PARSEC and MIST models yields a consistent age of $\sim4.9\pm 0.5$~Gyr. The photogeometric distance is $\langle d_{\rm pgeo} \rangle = 3325 \pm 634$~pc, with mean extinction $\langle A_{\rm V} \rangle = 0.383 \pm 0.125$~mag and color excess $E(B-V) = 0.124 \pm 0.040$~mag. The present Galactocentric radius is $R_{\rm GC} \approx 11.11$~kpc.

\item Spectroscopic catalog values from 138 members in the  GES~DR~5.1 catalog yield a mean metallicity of $\langle \rm [Fe/H] \rangle= -0.39\pm0.02$~dex. The abundance pattern shows a near-solar $\alpha$-element pattern and some enrichment in s-process elements, consistent with expectations for an older population, although detailed interpretations remain model-dependent.

\item The metallicity-corrected [Y/Mg]–age relation \citep{Casali2020} provides a chemical age of $4.73 \pm 2.39$~Gyr, which is consistent, within uncertainties, with the isochrone-based estimate. This agreement supports the adopted cluster age of $4.9\pm 0.5$~Gyr used throughout this study.

\item The photometric binary fraction of Berkeley~32 was estimated from the extinction-corrected CMD using the method of \citet{Milone2012} and \citet{Donada2023}. A total of 369 binary candidates were identified among 822 MS stars, yielding $f_{\rm b}=0.449\pm0.017$, for systems with mass ratios $q \geq 0.5$, which therefore represents a lower limit to the intrinsic binary fraction of the cluster.

\item Orbital integration with \texttt{galpy}, using both axisymmetric and non-axisymmetric potentials, suggests a moderately eccentric orbit ($e = 0.268 \pm 0.004$) with $R_{\rm peri} = 6.911 \pm 0.086$~kpc and $R_{\rm apo} = 11.971 \pm 0.059$~kpc. The maximum height ($Z_{\rm max} = 0.311 \pm 0.008$~kpc) is consistent with a dynamically heated thin-disk orbit. The guiding radius ($R_{\rm g} = 8.821 \pm 0.085$~kpc) differs from $R_{\rm GC}$ by $\sim2.3$~kpc, indicating a measurable level of epicyclic motion.

\item The chemical birth radius of $R_{\rm b} = 9.82$~kpc suggests an origin beyond the solar circle, and $\Delta R \approx -1$~kpc identifies Berkeley~32 as a moderate inward migrator, contrasting with the dominant outward trend among old OCs \citep{Otto2026}. The offsets $|R_{\rm g} - R_{\rm b}| \approx 1$~kpc and $|R_{\rm GC} - R_{\rm b}| \approx 1.3$~kpc, combined with $e = 0.268 \pm 0.004$, indicate contributions from both churning \citep{Netopil2022} and blurring, with its position in the age--radius deviation plane \citep{Frankel_2018, Frankel2020, Otto2026} falling below the empirical secular heating envelopes.

\item The dynamical relaxation time is estimated as $T_{\rm relax} = 30$~Myr, implying a large dynamical age parameter ($\tau \approx 164$) and an advanced state of dynamical evolution. Consistently, KS test results from the RCDFs reveal significant mass segregation: evolved stars are more centrally concentrated than MS stars, while MSTO stars occupy an intermediate distribution. This confirms that Berkeley~32 is a dynamically evolved OC in which internal relaxation processes have significantly influenced its present-day spatial structure.
\end{enumerate}

Overall, Berkeley~32 appears to be an old, metal-poor, and dynamically evolved OC that likely formed in the outer Galactic disk ($R_{\rm b} \approx 9.82$~kpc) and has since undergone orbital evolution. Its survival over $\sim5$~Gyr suggests a relatively stable dynamical history. The combined kinematic and chemical analysis provides valuable, albeit limited, insight into its evolution. Extending similar analyses to larger cluster samples with forthcoming spectroscopic surveys such as WEAVE and 4MOST will place these results on a firmer statistical basis.

\begin{acknowledgments}
We thank the anonymous referee for their constructive comments and suggestions, which helped improve the quality of the manuscript. Ing-Guey Jiang acknowledges support from the National Science and Technology Council (NSTC), Taiwan, under grants NSTC 113-2112-M-007-030 and NSTC 114-2112-M-007-029. This study is a part of the PhD Thesis of Deniz Cennet Çınar. This work uses data from the European Space Agency's Gaia mission, processed by the Gaia Data Processing and Analysis Consortium (DPAC). Based on data obtained from the ESO Science Archive Facility with DOI: https://doi.org/10.18727/archive/25
\end{acknowledgments}

\bibliography{references}{}
\bibliographystyle{aasjournalv7}

\appendix

\section{Gaia DR3 Data Retrieval and Quality Filtering}\label{app:data_quality}

The photometric uncertainties presented in Table~\ref{tab:photometric_errors} reflect the quality of the Gaia DR3 data after applying several selection criteria. Stars were retrieved from the Gaia DR3 archive using an ADQL query centered on the cluster, with a search radius of 18~arcmin and a source-brightness limit of $G = 21$~mag. The full query included positional constraints (\texttt{ra}, \texttt{dec}, and cone-search radius) and the magnitude limit; only the filter conditions are reproduced in the main text for brevity.

Before the membership analysis, we selected stars with a five-parameter astrometric solution (\texttt{astrometric\_params\_solved} $= 31$), ensuring that proper motion components and trigonometric parallaxes were available for all sources used in the analysis. These flagged sources were retained in the catalog and included in the membership analysis, but are noted as potentially less reliable for astrometry. Since membership probabilities are derived from the full proper-motion and parallax distributions, including these sources does not significantly bias the results; objects with discrepant astrometry naturally receive low membership probabilities. Overall, these filters and flags reduce and characterize the initial catalog, yielding a final sample of 50{,}837 sources toward Berkeley~32, as summarised in Table~\ref{tab:photometric_errors}. Before our analysis, we evaluated the photometric quality of the Gaia DR3 sample as a function of magnitude. As expected, photometric uncertainties remain very small for bright stars and increase progressively toward fainter magnitudes, with a more rapid rise beyond $G \approx 20$ mag. The detailed uncertainties in both $G$ magnitude and $(G_{\rm BP}-G_{\rm RP})$ color as a function of magnitude are summarized in Table~\ref{tab:photometric_errors}. Based on these trends, we adopt a conservative magnitude cut of $G = 20.1$~mag throughout this work to ensure reliable CMDs and minimize biases from stars with large photometric errors.

\begin{table}[htbp]
\centering
\caption{Mean photometric uncertainties in Gaia $G$ magnitudes and $(G_{\rm BP}-G_{\rm RP})$ colors for stars toward Berkeley~32. Values are presented as a function of $G$ magnitude intervals, showing the typical precision achieved in each bin.}
\label{tab:photometric_errors}
\begin{tabular}{c ccc}
\toprule
$G$ (mag) & $N$ & $\sigma_G$ & $\sigma_{G_{\rm BP}-G_{\rm RP}}$\\
\midrule
 6--14 &  1085 & 0.003 & 0.006 \\
14--15 &  1231 & 0.003 & 0.006 \\
15--16 &  2425 & 0.003 & 0.007 \\
16--17 &  4436 & 0.003 & 0.011 \\
17--18 &  6843 & 0.003 & 0.023 \\
18--19 &  9416 & 0.004 & 0.053 \\
19--20 & 11739 & 0.006 & 0.121 \\
20--21 & 13303 & 0.013 & 0.271 \\
21--23 &   359 & 0.031 & 0.395 \\
\midrule
Total  & 50837 & 0.007 & 0.116 \\
\bottomrule
\end{tabular}
\end{table}

\section{Blue Straggler and Variable Star Catalog} \label{sec:bss_variable}

The complete catalog of BSS, YSS, and variable star candidates identified in Section~\ref{sec:stellar_content} is provided in Table~\ref{tab:appendix_bss}. The table lists each star alongside its equatorial coordinates, $G$-band magnitude, membership probability, and primary classification (e.g., BSS or YSS). Detailed cross-match references and variability types are indicated in the table using superscripts. The superscript $^{a}$ denotes the origin of the data: (1) corresponds to newly identified sources in this study, (2) to \citet{rain2021new}, and (3) to \citet{Jadhav2019blue}. Variable star classifications encompass eclipsing binaries (ECL), RS Canum Venaticorum-type chromospherically active binaries (RS), and various pulsating variables, namely $\delta$~Scuti (DSCT), $\gamma$~Doradus (GDOR), and SX~Phoenicis (SXPHE).

\begin{sidewaystable*}
\caption{Full list of BSS, YSS, and variable star candidates along with their membership probabilities. Superscript $^{a}$ denotes the data origin and variability classification, respectively (see Appendix~\ref{sec:bss_variable} for detailed definitions).} \label{tab:appendix_bss}
\begin{longtable}{l l l c c l l  | l l l c c l l}
\toprule
No. & $\alpha$ (hh:mm:ss) & $\delta$ (dd:mm:ss) & $G$ (mag) & $P$ & Type & Ref.$^{a}$ &
No. & $\alpha$ (hh:mm:ss) & $\delta$ (dd:mm:ss) & $G$ (mag) & $P$ & Type & Ref.$^{a}$ \\
\midrule
\endfirsthead
\multicolumn{14}{c}{\tablename\ \thetable{} -- \textit{Continued}} \\
\toprule
No. & $\alpha$ (hh:mm:ss) & $\delta$ (dd:mm:ss) & $G$ (mag) & $P$ & Type & Ref.$^{a}$ &
No. & $\alpha$ (hh:mm:ss) & $\delta$ (dd:mm:ss) & $G$ (mag) & $P$ & Type & Ref.$^{a}$ \\
\midrule
\endhead
\midrule
\endfoot
\bottomrule
 1 & 06:56:50.81 & $+$06:26:49.8 & 15.53 & 1.00 & BSS           & 1     & 39 & 06:59:24.39 & $+$06:30:50.0 & 15.12 & 1.00 & BSS       & 1     \\
 2 & 06:57:02.04 & $+$06:21:55.2 & 14.46 & 1.00 & BSS, DSCT& 1    & 40 & 06:59:26.43 & $+$06:24:49.6 & 15.22 & 1.00 & BSS       & 1     \\
 3 & 06:57:20.65 & $+$06:29:00.4 & 14.90 & 0.99 & BSS, DSCT& 1    & 41 & 06:58:08.16 & $+$06:25:24.2 & 14.53 & 1.00 & BSS       & 1,2,3 \\
 4 & 06:57:20.69 & $+$06:26:45.2 & 14.10 & 1.00 & BSS           & 1     & 42 & 06:58:09.31 & $+$06:24:37.8 & 15.48 & 1.00 & BSS       & 1,2,3 \\
 5 & 06:57:21.63 & $+$06:34:33.0 & 15.73 & 1.00 & BSS           & 1     & 43 & 06:58:11.00 & $+$06:26:06.9 & 16.13 & 1.00 & BSS       & 1,2,3 \\
 6 & 06:57:23.31 & $+$06:25:16.4 & 15.68 & 1.00 & BSS           & 2,3   & 44 & 06:58:11.01 & $+$06:08:08.9 & 15.37 & 1.00 & BSS       & 1     \\
 7 & 06:57:24.11 & $+$06:09:60.0 & 14.24 & 0.75 & BSS, DSCT& 1    & 45 & 06:58:12.60 & $+$06:26:30.2 & 15.85 & 1.00 & BSS       & 1,2   \\
 8 & 06:57:29.98 & $+$06:12:57.7 & 14.03 & 0.86 & BSS           & 1     & 46 & 06:58:13.27 & $+$06:31:00.7 & 15.50 & 1.00 & BSS, DSCT       & 1     \\
 9 & 06:57:34.63 & $+$06:16:47.4 & 14.12 & 1.00 & BSS           & 1     & 47 & 06:58:13.31 & $+$06:35:49.0 & 15.48 & 1.00 & BSS       & 1     \\
10 & 06:57:35.92 & $+$06:14:11.5 & 14.90 & 1.00 & BSS           & 1     & 48 & 06:58:13.96 & $+$06:09:59.6 & 15.72 & 0.82 & BSS       & 1     \\
11 & 06:57:37.43 & $+$06:13:01.4 & 14.73 & 0.98 & BSS           & 1     & 49 & 06:58:14.33 & $+$06:22:57.6 & 16.09 & 1.00 & BSS       & 1,2,3 \\
12 & 06:57:37.65 & $+$06:43:56.6 & 15.26 & 1.00 & BSS           & 1     & 50 & 06:58:16.78 & $+$06:22:05.6 & 15.57 & 1.00 & BSS       & 1,2,3 \\
13 & 06:57:39.26 & $+$06:35:50.9 & 16.03 & 0.99 & BSS           & 2,3   & 51 & 06:58:18.31 & $+$06:30:41.5 & 13.46 & 1.00 & BSS       & 1,2,3 \\
14 & 06:57:41.09 & $+$06:38:41.0 & 15.10 & 1.00 & BSS           & 1     & 52 & 06:58:19.69 & $+$06:37:17.3 & 15.94 & 0.97 & BSS       & 1     \\
15 & 06:57:41.68 & $+$06:41:55.2 & 15.25 & 0.98 & BSS           & 1     & 53 & 06:58:19.92 & $+$06:21:45.9 & 15.57 & 1.00 & BSS       & 1     \\
16 & 06:57:45.20 & $+$06:15:10.1 & 14.65 & 0.99 & BSS           & 2,3   & 54 & 06:58:21.19 & $+$06:27:27.5 & 15.11 & 0.83 & BSS       & 1,2   \\
17 & 06:57:45.98 & $+$06:43:51.0 & 15.60 & 0.98 & BSS           & 1     & 55 & 06:58:21.63 & $+$06:29:47.3 & 14.90 & 1.00 & BSS       & 1     \\
18 & 06:57:46.08 & $+$06:26:10.8 & 15.77 & 1.00 & BSS           & 2,3   & 56 & 06:58:22.19 & $+$06:24:02.8 & 15.65 & 0.98 & BSS       & 1,2,3 \\
19 & 06:57:47.06 & $+$06:25:11.6 & 16.37 & 0.94 & BSS           & 2     & 57 & 06:58:22.25 & $+$06:27:00.8 & 15.39 & 1.00 & BSS       & 1     \\
20 & 06:57:47.99 & $+$06:41:36.3 & 14.53 & 0.86 & BSS           & 1     & 58 & 06:58:23.83 & $+$06:14:47.4 & 14.25 & 0.99 & BSS, DSCT       & 1     \\
21 & 06:57:48.46 & $+$06:17:13.5 & 15.67 & 0.92 & BSS           & 1     & 59 & 06:58:29.44 & $+$06:22:05.4 & 16.18 & 1.00 & BSS       & 1,2,3 \\
22 & 06:57:48.91 & $+$06:17:31.2 & 14.73 & 1.00 & BSS           & 1     & 60 & 06:58:30.67 & $+$06:33:38.4 & 15.45 & 1.00 & BSS       & 1,2,3 \\
23 & 06:57:49.33 & $+$06:43:54.8 & 15.67 & 1.00 & BSS           & 1     & 61 & 06:58:35.93 & $+$06:29:46.2 & 14.56 & 1.00 & BSS       & 1     \\
24 & 06:57:50.55 & $+$06:13:16.0 & 14.46 & 0.91 & BSS           & 1     & 62 & 06:58:38.89 & $+$06:40:35.4 & 14.28 & 1.00 & BSS       & 1     \\
25 & 06:57:51.20 & $+$06:28:07.0 & 15.76 & 1.00 & BSS           & 2,3   & 63 & 06:58:39.85 & $+$06:34:16.1 & 15.32 & 0.96 & BSS, DSCT       & 1     \\
26 & 06:57:56.11 & $+$06:16:43.9 & 16.63 & 1.00 & BSS           & 2     & 64 & 06:58:41.76 & $+$06:15:54.0 & 15.11 & 0.95 & BSS       & 1     \\
27 & 06:57:57.36 & $+$06:23:40.6 & 15.52 & 1.00 & BSS           & 1     & 65 & 06:58:44.79 & $+$06:27:15.9 & 14.82 & 1.00 & BSS       & 1     \\
28 & 06:57:58.50 & $+$06:17:46.1 & 15.35 & 1.00 & BSS           & 1     & 66 & 06:58:58.10 & $+$06:24:43.7 & 16.04 & 1.00 & BSS       & 1     \\
29 & 06:57:59.42 & $+$06:27:36.8 & 15.57 & 1.00 & BSS           & 1     & 67 & 06:59:05.67 & $+$06:15:01.7 & 15.69 & 1.00 & BSS       & 1     \\
30 & 06:57:59.94 & $+$06:29:00.5 & 16.05 & 0.97 & BSS           & 2     & 68 & 06:59:08.53 & $+$06:35:03.5 & 11.64 & 1.00 & BSS       & 1     \\
31 & 06:58:01.33 & $+$06:43:49.2 & 15.29 & 0.74 & BSS           & 1     & 69 & 06:59:12.34 & $+$06:17:29.9 & 14.75 & 1.00 & BSS       & 1     \\
32 & 06:58:03.94 & $+$06:29:07.8 & 15.76 & 0.76 & BSS           & 1     & 70 & 06:59:13.67 & $+$06:21:30.3 & 15.51 & 1.00 & BSS       & 1     \\
33 & 06:58:04.97 & $+$06:09:48.3 & 15.65 & 1.00 & BSS           & 1     & 71 & 06:58:08.13 & $+$06:25:35.5 & 14.40 & 1.00 & YSS       & 1,2   \\
34 & 06:58:05.05 & $+$06:24:59.3 & 15.84 & 1.00 & BSS           & 2,3   & 72 & 06:58:12.58 & $+$06:25:44.8 & 14.90 & 1.00 & YSS       & 1,2   \\
35 & 06:58:05.09 & $+$06:25:10.4 & 15.45 & 1.00 & BSS           & 1     & 73 & 06:58:03.72 & $+$06:26:18.6 & 15.97 & 1.00 & RS       & 1   \\ 
36 & 06:58:07.09 & $+$06:18:24.4 & 16.01 & 1.00 & BSS           & 2,3   & 74 & 06:57:17.53 & $+$06:23:29.5 & 14.64 & 0.81 & RS           & 1    \\
37 & 06:58:07.82 & $+$06:25:57.3 & 15.65 & 1.00 & BSS, ECL& 1,2,3 & 75 & 06:58:37.21 & $+$06:35:58.6 & 16.48 & 0.96 & DSCT      & 1     \\
38 & 06:58:07.86 & $+$06:24:46.9 & 15.64 & 1.00 & BSS           & 1     & 76 & 06:59:08.02 & $+$06:24:48.6 & 16.11 & 1.00 & DSCT     & 1     \\
\end{longtable}

\end{sidewaystable*}

\section{Luminosity and Mass Functions}
\label{sec:mf}
\begin{figure*}[htbp]
\centering
\includegraphics[width=0.45\linewidth]{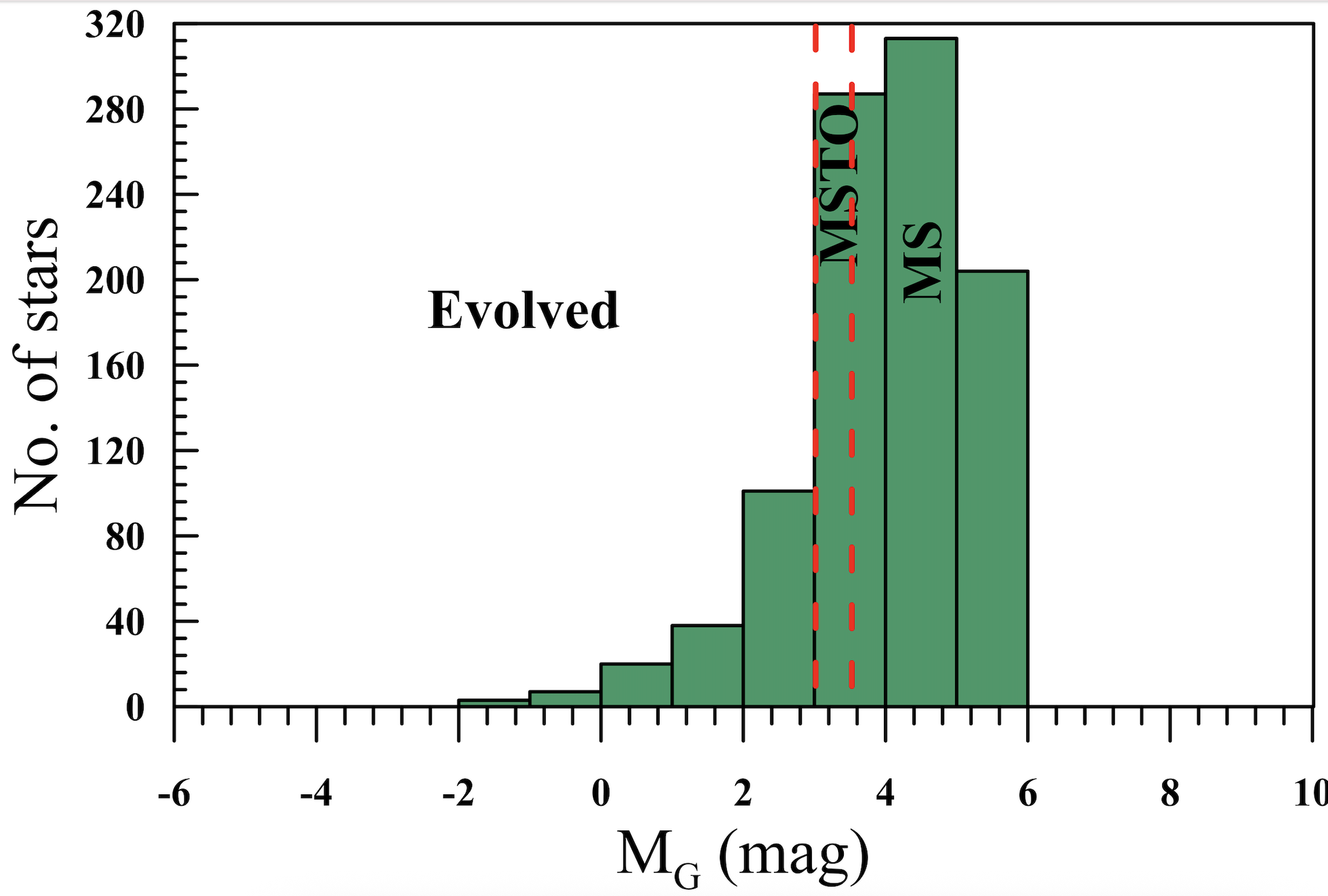}
\caption{Luminosity function (LF) of Berkeley~32 in the absolute $G$ band, showing the number of probable members as a function of magnitude.}
\label{LFs}
\end{figure*}

The luminosity function (LF) describes the distribution of cluster members as a function of absolute magnitude. For Berkeley~32, the $G$-band absolute magnitude range was divided into equal bins, and the number of probable members in each bin was counted. Absolute magnitudes were derived from apparent magnitudes using the distance modulus obtained from the isochrone analysis. The \textit{Gaia} DR3 photometric catalogue is effectively complete over the magnitude range considered in this study. To minimize any potential effects of photometric incompleteness, the luminosity and mass functions were derived only for high-probability cluster members brighter than $G=19$~mag, well above the faint limit of the \textit{Gaia} DR3 catalogue. Consequently, the adopted sample is not expected to be significantly affected by photometric incompleteness, and the derived luminosity and mass functions are therefore considered robust over the investigated magnitude range. The resulting LF is shown in Figure~\ref{LFs}.

To estimate stellar masses, we used PARSEC isochrones \citep{Bressan2012} at the adopted cluster age to construct the mass--luminosity relation (MLR), which was fitted with a fourth-degree polynomial:
\begin{equation}
M_{\rm c} = a_0 + a_1 M_{\rm G} + a_2 M_{\rm G}^2 + a_3 M_{\rm G}^3 + a_4 M_{\rm G}^4.
\label{Eq:ML_new}
\end{equation}
The derived fit coefficients for Berkeley~32 are $a_0 = 1.1848$, $a_1 = 0.0170$, $a_2 = 0.006$, $a_3 = -0.005$, and $a_4 = 0.0003$ \citep{2017SerAJ.194...59A}. This fit covers the magnitude range of $-1.356$ to $5.962$~mag. Using this relation, we estimated a mean stellar mass of $\langle M_{\rm C} \rangle = 1.00~M_\odot$ and a total cluster mass of $M_{\rm C} \approx 1125~M_\odot$. Deriving a reliable mass function slope for Berkeley~32 is not feasible. As an old cluster with a turn-off mass of $\sim1.3~M_\odot$, the observable mass range is narrow, which causes the mass function to appear artificially flat, consistent with previous work \citep[e.g.,][]{Sariya2021}. We therefore treat the derived total mass of $1125~M_\odot$ as a lower limit and do not attempt to characterize the global mass distribution.

\section{TESS Variability Analysis}
\label{app:tess}

The \textit{Transiting Exoplanet Survey Satellite} \citep[\textit{TESS};][]{ricker2015transiting} uses four $2\mathrm{K}\times2\mathrm{K}$ CCD cameras with a total field of view of $24^{\circ}\times96^{\circ}$ and a pixel scale of about ${\sim}21^{\prime\prime}\,\mathrm{pixel}^{-1}$. \textit{TESS} observes the sky in sectors lasting about 27 days and performs time-series photometry in the 600--1000 nm wavelength range. During the primary mission (Sectors~1--26), full-frame images (FFIs) were obtained every 30 minutes. In the first extended mission (Sectors~27--55), the cadence improved to 10 minutes. From Sector~56 onward, 200~s target pixel files (TPFs) became available for selected targets. All data products were processed by the \textit{TESS} Science Processing Operations Center (SPOC) pipeline\footnote{\url{https://heasarc.gsfc.nasa.gov/docs/tess/documentation.html}}. We searched the Mikulski Archive for Space Telescopes \citep[MAST;][]{tess_mast_2021} for publicly available \textit{TESS} observations in the direction of Berkeley~32. Light curves and TPFs were downloaded and analysed using the \textsc{lightkurve} package \citep{lightkurve2018}.

For each light curve, we calculated the normalised flux and computed a Lomb--Scargle (LS) periodogram \citep{lomb1976, scargle1982} using the \textsc{astropy} implementation \citep{robitaille2013astropy}. False-alarm probability (FAP) levels of 10\%, 5\%, and 1\% were estimated with bootstrap resampling. The period with the highest LS power above the 99\% confidence level was adopted as the best-period candidate, and the light curve was phase-folded using this period.

\begin{figure}[htbp]
    \centering
    \includegraphics[width=1\linewidth]{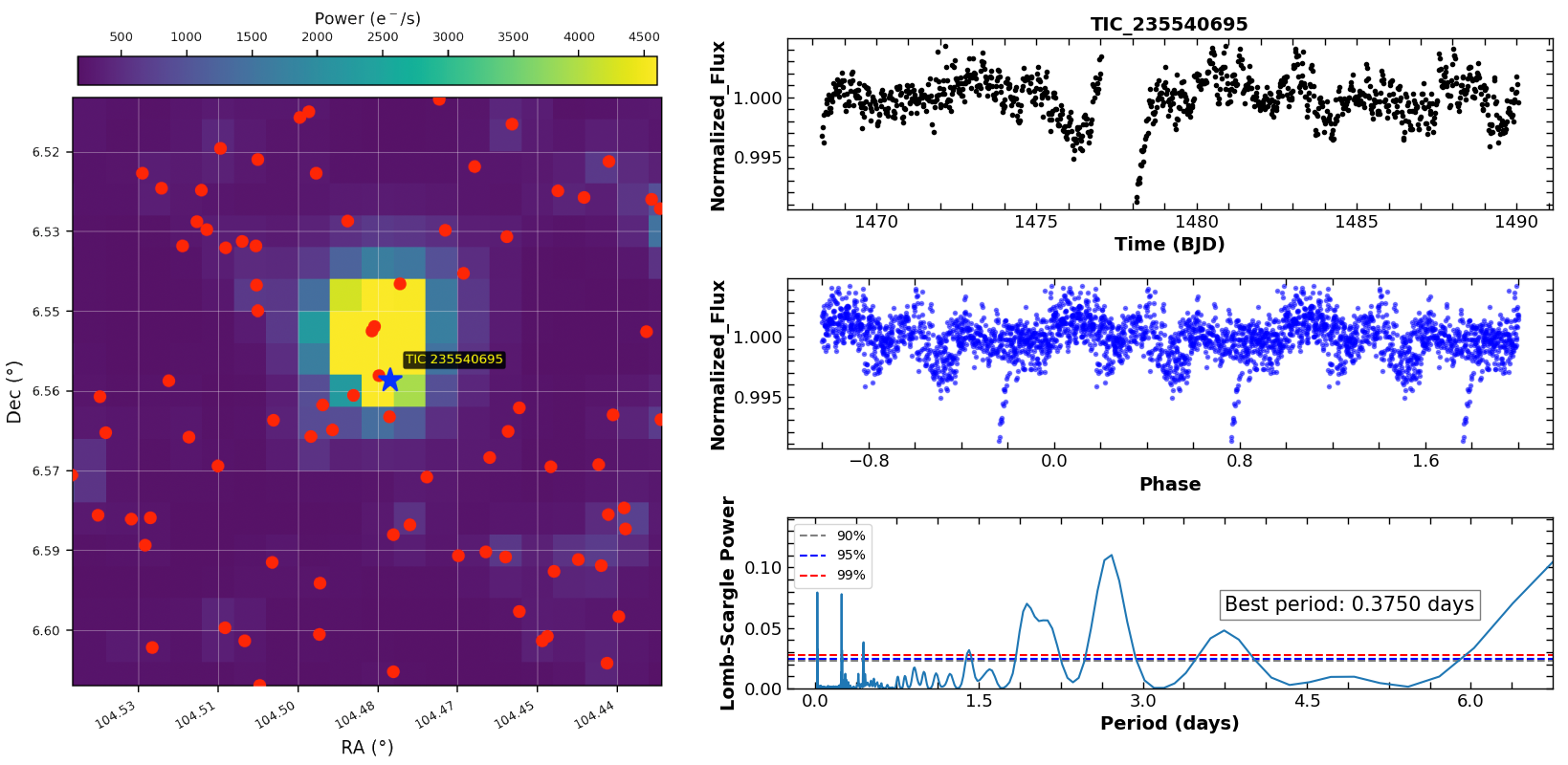}
    \caption{\textit{Left:} \textit{TESS} pixel-level finder chart for TIC~235540695, showing the TESScut FFI cutout with the median flux (power) map. The yellow star marks the target, and red circles indicate nearby TIC sources. \textit{Top right:} Normalised flux versus time (BJD\,$-$\,2457000). \textit{Middle right:} Phase-folded light curve at the best-fit period of $P=0.3750$~d. \textit{Bottom right:} LS periodogram with 90\% (grey dashed), 95\% (blue dashed), and 99\% (red dashed) false-alarm probability levels.}
    \label{fig:tess_example}
\end{figure}

An example is shown in Figure~\ref{fig:tess_example}. The top panel shows the raw \textit{TESS} light curve of TIC~235540695. The middle panel presents the phase-folded light curve using the best-fit period of $P=0.3750$~d. The bottom panel shows the LS periodogram together with the 90\%, 95\%, and 99\% FAP levels. The strongest peak appears near $P\approx1.8$~d and is above the 99\% confidence level, while the adopted period of 0.3750~d is likely a harmonic of this signal. However, the large \textit{TESS} pixel scale (${\sim}21^{\prime\prime}\,\mathrm{pixel}^{-1}$) causes strong crowding and flux contamination in the cluster field, making it difficult to identify the origin of variability for individual stars. In addition, the limited number of observed sectors and gaps caused by downlink windows and scattered light reduce the temporal coverage needed for reliable period analysis. Because of these limitations, a detailed variability study of Berkeley~32 members using only \textit{TESS} data is beyond the scope of this work. The example shown here is included only to illustrate the quality of the available \textit{TESS} data for this cluster field.

\end{document}